\documentclass[amssymb,nofootinbib,nobibnotes,aps,prd,preprintnumbers]{revtex4}

 \usepackage{here}

\usepackage[dvipdfmx]{graphicx}
\usepackage{amssymb,amsfonts,amsmath,bm,color,multirow,cases,empheq,hyperref}
\usepackage{mathrsfs}

\definecolor{darkgreen}{rgb}{0,0.5,0.3}

\usepackage{enumerate}
\usepackage{ulem}
\usepackage{afterpage}

\hypersetup{%
 setpagesize=false,
 bookmarksnumbered=true,%
 bookmarksopen=true,%
 colorlinks=true,%
 linkcolor=blue,%
 citecolor=red}

\newcommand{\fr}[1]{\frac{1}{#1}}

\newcommand{\nonum}{\nonumber\\ }

\newcommand{\cout}[1]{}

\newcommand{\arrayL}[1]{\left(\begin{array}{#1}}
\newcommand{\arrayR}{\end{array}\right)}
\newcommand{\arrayLb}[1]{\left[\begin{array}{#1}}
\newcommand{\arrayRb}{\end{array}\right]}

\begin{document}

\title{New construction of a charged dipole black ring by Harrison transformation}

\author{Ryotaku Suzuki}
\email{sryotaku@toyota-ti.ac.jp}
\author{Shinya Tomizawa}
\email{tomizawa@toyota-ti.ac.jp}
\affiliation{\vspace{3mm}Mathematical Physics Laboratory, Toyota Technological Institute\vspace{2mm}\\Hisakata 2-12-1, Tempaku-ku, Nagoya, Japan 468-8511\vspace{3mm}}

\begin{abstract}

We present  an exact solution for a non-BPS charged rotating black ring endowed with a dipole charge in the bosonic sector of five-dimensional minimal supergravity. 
Utilizing the electric Harrison transformation, we derive this solution by converting a five-dimensional vacuum solution into a charged solution within the realm of five-dimensional minimal supergravity.
As the seed solution for the Harrison transformation,  we use a vacuum solution of a rotating black ring possessing a Dirac-Misner string singularity.  The resulting solution exhibits regularity, indicating the absence of curvature singularities, conical singularities, orbifold singularities,  Dirac-Misner string singularities, and closed timelike curves both on and outside the horizon. 
This obtained solution carries the mass, two angular momenta, an electric charge, and a dipole charge, with only three of these quantities being independent, similar to the charged rotating dipole black ring found previously by Elvang, Emparan and Figueras.
However, aside from the vacuum case, these two solutions do not coincide.
We discuss the difference between them in the phase space.

\end{abstract}

\date{\today}
\preprint{TTI-MATHPHYS-25}

\maketitle

\section{Introduction}

In the realm of string theory and related fields, higher-dimensional black holes and other extended black objects have played a significant role in understanding such higher-dimensional theories over the past two decades~\cite{Emparan:2008eg,Emparan:2006mm}.
In particular, the physics of black holes in five-dimensional minimal supergravity (Einstein-Maxwell-Chern-Simons theory) has garnered increased attention, as it serves as a low-energy limit of string theory.
This five-dimensional minimal supergravity bears resemblance to eleven-dimensional supergravity in terms of its Lagrangians, with  a three-form field being replaced by Maxwell's $U(1)$ gauge field.
The similarity between five-dimensional minimal supergravity and eleven-dimensional supergravity has been studied previously~\cite{Mizoguchi:1998wv,Mizoguchi:1999fu}.
Furthermore, five-dimensional supergravity can be obtained through a truncated toroidal compactification of eleven-dimensional supergravity by identifying three vector fields and freezing out the moduli~\cite{Cremmer:1997ct,Cremmer:1998px}.
This underscores the importance of finding all exact solutions of black holes in five-dimensional minimal supergravity and classifying them, as it contributes to our understanding of string theory.
However, this has not yet been achieved, although various exact solutions of black holes in this theory have been generated with the help of recent developments in solution-generation techniques~\cite{Bouchareb:2007ax,Ford:2007th,Galtsov:2008pbf,Galtsov:2008bmt,Galtsov:2008jjb,Compere:2009zh,Figueras:2009mc,Mizoguchi:2011zj}.

\medskip

The uniqueness theorem for charged rotating black holes in five-dimensional minimal supergravity~\cite{Tomizawa:2009ua} says that, assuming the presence of two commuting axial isometries, spherical topology of the horizon cross-section, and trivial topology (${\Bbb R} \times \{ {\Bbb R}^4 \setminus {\Bbb B}^4 \}$) of the domain of outer communication, an asymptotically flat, stationary charged rotating black hole with a non-degenerate horizon is uniquely specified by its mass, charge, and two independent angular momenta, thus being described by the five-dimensional Cveti{\v c}-Youm solution. 
Similarly, according to the uniqueness theorem for charged black rings in the same supergravity~\cite{Tomizawa:2009tb}, 
assuming the same Killing symmetries,  trivial topology (${\Bbb R} \times \{ {\Bbb R}^4 \setminus  D^2\times S^2 \} $) 
of the domain of outer communication, an asymptotically flat, stationary charged rotating black ring with non-degenerate connected event horizon of cross-section topology $S^1\times S^2$ --- if this exists --- is characterized uniquely by the mass, electric charge, two independent angular momenta, dipole charge, and additional information on the rod structure such as the ratio of the $S^2$ radius to the $S^1$ radius.
When a rotating black ring interacts with the Maxwell field, it induces a type of dipole charge. 
Consequently, the dipole charge, which is not a conserved charge, serves as an additional parameter to characterize the black ring. 
This was illustrated in the first example of the dipole black ring solution found by Emparan~\cite{Emparan:2004wy}, which is electrically coupled to a two-form or a dual magnetic one-form field. 
Further examples of dipole rings were constructed by Elvang {\it et al.}~\cite{Elvang:2004xi} in five-dimensional minimal supergravity, originating from a seven-parameter family of non-supersymmetric black ring solutions.
However, this dipole black ring solution does not admit a limit to a supersymmetric solution, and moreover, the dipole charge of their solution is not an independent parameter since it is related to the other conserved charges. 
More precisely, the dipole black ring solution has four conserved charges and a dipole charge, of which only three quantities are independent. 
As conjectured by the authors in Ref.~\cite{Elvang:2004xi}, 
it is anticipated that a more general non-Bogomol’nyi-Prasad-Sommerfield (BPS) black ring solution exists, characterized by its mass, two independent angular momenta, electric charge, and a dipole charge, which are independent of the other asymptotic conserved charges. 
Following that,  Feldman and Pomeransky~\cite{Feldman:2014wxa} seem to have made a significant advancement by presenting the most general black ring solution with mass, two angular momenta, three electric charges and three dipole charges in five-dimensional $U(1)^3$ supergravity, which also encompasses the most general black ring solution with the five independent quantities in the five-dimensional minimal supergravity.

\medskip

Dimensionally reduced gravity theories and supergravity possess a global symmetry  known as ``hidden symmetry," which often proves to be a powerful tool in discovering new solutions. 
New solutions can be obtained by applying this group transformation to a known solution within the same theory, which is referred to as a ``seed solution" (see Ref.\cite{Ehlers,Harrison,exact} for four-dimensional Einstein gravity). 
The dimensional reduction of five-dimensional minimal supergravity to four dimensions was studied in Refs.~\cite{Chamseddine:1980mpx,Mizoguchi:1998wv}, and the reduced theory precisely exhibits $SL(2, {\Bbb R})$ symmetry, obtained by the dimensional reduction of eleven-supergravity~\cite{Cremmer:1979up}. 
The new solution-generation technique using this $SL(2,{\Bbb R})$ symmetry~\cite{Mizoguchi:2011zj} actually generated the Kaluza-Klein black hole solutions~\cite{Mizoguchi:2012vg,Tomizawa:2012nk}.
As first studied by Mizoguchi and Ohta~\cite{Mizoguchi:1998wv,Mizoguchi:1999fu} in five-dimensional minimal supergravity, the existence of two commuting Killing vector fields reduces the theory to a three-dimensional non-linear sigma model with a $G_{2(2)}$ target space symmetry. 
In the presence of two spacelike commuting Killing vector fields, it is described by the $G_{2(2)}/SO(4)$ sigma model coupled to gravity, and if one of the two commuting Killing vector fields is timelike, the symmetry is replaced with $G_{2(2)}/[SL(2,{\Bbb R})\times SL(2,{\Bbb R})]$.
Using this $G_{2(2)}$ symmetry, Bouchareb et al.~\cite{Bouchareb:2007ax} developed a solution-generation technique including an electric Harrison transformation, which transforms a five-dimensional vacuum solution into an electrically charged solution in five-dimensional minimal supergravity. By constructing the representation of a coset in terms of a $7\times 7$ matrix, the application of the transformation to the five-dimensional vacuum rotating black hole (the Myers-Perry solution~\cite{Myers:1986un}) yields the five-dimensional charged rotating black hole (the Cveti\v{c}-Youm solution~\cite{Cvetic:1996xz}). However, the transformation of the vacuum doubly rotating black ring (the Pomeransky-Sen'kov solution~\cite{Pomeransky:2006bd}) does not generate a regular charged doubly spinning black ring solution, as the resulting solution suffers from an inevitable Dirac-Misner string singularity.

 \medskip

 The goal of this paper is to solve the undesirable and critical problem of the inevitable appearance of the Dirac-Misner string singularity after the Harrison transformation and to construct a charged rotating black ring solution with a dipole charge in the $C$-metric form in five-dimensional minimal supergravity.  
 The basic procedure is as follows: First, using the inverse scattering method (ISM)~\cite{Belinsky:1979mh,Belinski:2001ph,Pomeransky:2005sj} for five-dimensional Einstein gravity, we construct a vacuum solution of a rotating black ring possessing a Dirac-Misner string singularity inside the ring. 
 Secondly, by performing the electric Harrison transformation for this vacuum solution, we obtain a charged rotating black ring solution with a Dirac-Misner string singularity. 
Finally,  to ensure the regularity of the obtained solution, we choose appropriate parameters such that the Dirac-Misner string singularity inside the black ring disappears. 
 Consequently, the final resulting solution describes a charged rotating dipole black ring, which is regular in the sense that it lacks curvature singularities, conical singularities, Dirac-Misner string singularities, or orbifold singularities on and outside the horizon, and is also free from closed timelike curves (CTCs). 
 This fundamental approach is reminiscent of the solitonic construction of the $S^1$-rotating black ring, where a singular seed solution was chosen to generate the regular solution~\cite{Iguchi:2006rd,Tomizawa:2006vp,Emparan:2008eg}. 
 Similar to the charged rotating black ring solution~\cite{Elvang:2004xi}
 obtained earlier by Elvang, Emparan, and Figueras (EEF), the obtained black ring possesses five physical quantities: its mass, two angular momenta, an electric charge, and a dipole charge, of which only three are independent. However, it differs from the EEF black ring in that the former has two non-zero horizon angular velocities, whereas the latter has only one, despite both having two non-zero angular momenta (this discrepancy arises because the rotating Maxwell field outside the horizon contributes to the angular momentum corresponding to the zero horizon angular velocity).

\medskip

The remainder of the paper is dedicated to constructing the aforementioned  black ring solution. 
In the following section~\ref{sec:setup}, we outline the setup and formulation. 
Then, in Sec.~\ref{sec:ism}, we construct a vacuum solution of a rotating black ring with a Dirac-Misner string singularity between the horizon and the ring center as the seed solution for the Harrison transformation. 
At this stage, we retain the Dirac-Misner string singularity without attempting to eliminate it. 
Next, in Sec.~\ref{sec:sol}, we apply the electric Harrison transformation to the vacuum solution, resulting in a corresponding charged solution within five-dimensional minimal supergravity. 
By imposing boundary conditions on the parameters, we derive a  charged, rotating dipole black ring solution in the $C$-metric form, demonstrating its absence of singularities (including curvature singularity, conical singularity, and Dirac-Misner string singularity) as well as closed timelike curves (CTCs) both on and outside the horizon. 
In Sec.~\ref{sec:phase}, we delve into the phase of the obtained solution, followed by a comparison with the  EEF black ring solution in Sec.~\ref{sec:EEF}. 
Finally, in Sec.~\ref{sec:sum}, we summarize our findings.

\section{setup}\label{sec:setup}

Let us begin with a basic setup for  asymptotically flat, stationary and bi-axisymmetric solutions in the bosonic sector of the five-dimensional minimal ungauged supergravity (Einstein-Maxwell-Chern-Simons theory), whose action takes the form
\begin{eqnarray}
S=\frac{1}{16 \pi G_5}  \left[ 
        \int d^5x \sqrt{-g}\left(R-\frac{1}{4}F^2\right) 
       -\frac{1}{3\sqrt{3}} \int F\wedge F\wedge A 
  \right] \,, 
\label{action} 
\end{eqnarray} 
where $F=dA$. The field equation consists of the Einstein equation 
\begin{eqnarray}
 R_{\mu \nu } -\frac{1}{2} R g_{\mu \nu } 
 = \frac{1}{2} \left( F_{\mu \lambda } F_\nu^{ ~ \lambda } 
  - \frac{1}{4} g_{\mu \nu } F_{\rho \sigma } F^{\rho \sigma } \right) \,, 
 \label{Eineq}
\end{eqnarray}
and the Maxwell equation with a Chern-Simons term
\begin{eqnarray}
 d\star F+\frac{1}{\sqrt{3}}F\wedge F=0 \,. 
\label{Maxeq}
\end{eqnarray}
Assuming the existence of one timelike Killing vector $\xi_0 = \partial/\partial t$ and one spacelike axial Killing vector $\xi_1=\partial/\partial\psi$, this theory reduces to the $G_{2(2)}/SL(2,{\Bbb R})\times SL(2,{\Bbb R})$ non-linear sigma models coupled to three-dimensional gravity~\cite{Mizoguchi:1998wv,Mizoguchi:1999fu}.
Under a further assumption of the presence of the third spacelike axial Killing vector $\xi_2=\partial/\partial\phi$, i.e., the existence  of three mutually commuting Killing vectors, the metric can be written in the Weyl-Papapetrou form\footnote{If one choose two axial Killing vectors for the reduction, then $\lambda_{ab}$ has the Riemanian signature and one has to flip the sign of the line elements $d\rho^2+dz^2$ as in Refs.~\cite{Tomizawa:2009ua,Tomizawa:2009tb}. }
\begin{eqnarray}
ds^2
&=& \lambda_{ab}(dx^a+a_\phi^a{}d\phi)(dx^b+a_\phi^b{}d\phi)+\tau^{-1}\rho^2 d\phi^2
     +\tau^{-1}e^{2\sigma}(d\rho^2+dz^2) \,, \label{eq:WPform}
\end{eqnarray}
and the gauge potential is written, 
\begin{eqnarray} 
 A = \sqrt{3}\psi_a dx^a + A_\phi d\phi \,, 
\label{pote:gauge}
\end{eqnarray}
where the coordinates $x^a=(t,\psi)$ ($a=0,1$) denote the Killing coordinates, and thus all functions $\lambda_{ab}$, 
$\tau:=-{\rm det}(\lambda_{ab})$, $a^a$, $\sigma$, and $(\psi_a,A_\phi)$ are independent of $\phi$ and $x^a$. 
Note that the coordinates $(\rho,z)$ that span a two-dimensional base space, $\Sigma=\{(\rho,z)|\rho\ge 0,\ -\infty<z<\infty \}$, 
are globally well-defined, harmonic, and mutually conjugate on $\Sigma$. 

\medskip
As discussed in Ref.~\cite{Tomizawa:2009ua}, by using Eqs.~(\ref{Eineq}) and (\ref{Maxeq}), we can introduce the magnetic potential $\mu$ and twist potentials $\omega_a$ by   
\begin{eqnarray} 
d\mu&=&\frac{1}{\sqrt{3}}\star (\xi_0\wedge \xi_1\wedge F) 
    - \epsilon^{ab}\psi_ad\psi_b \,, 
\label{eq:mu} 
\\
 d\omega_a&=&\star (\xi_0\wedge \xi_1 \wedge d\xi_a)+\psi_a(3d\mu+\epsilon^{bc}\psi_bd\psi_c)\,, 
\label{eq:twistpotential} 
\end{eqnarray} 
where $\epsilon^{01}=-\epsilon^{10}=1$.
The metric functions $a_{\phi}^a\ (a=0,1)$ and the component $A_\phi$ of the gauge potential are determined by the eight scalar functions $\{\lambda_{ab},\omega_a,\psi_a,\mu\}$ from Eqs.~(\ref{eq:mu}) and (\ref{eq:twistpotential}), and the function $\sigma$ is also determined by these scalar functions up to a constant factor. 
Then, the action~(\ref{action}) reduces to the nonlinear sigma model  for the eight scalar functions $\{\lambda_{ab},\omega_a,\psi_a,\mu\}$ invariant under the $G_{2(2)}$-transformation. 

\medskip
In particular, utilizing the $G_{2(2)}$ symmetry, Ref.~\cite{Bouchareb:2007ax} constructed the electric Harrison transformation preserving asymptotic flatness that transform a five-dimensional vacuum solution $\{\lambda_{ab},\omega_a,\psi_a=0,\mu=0\}$ into a charged solution $\{\lambda'_{ab},\omega'_a,\psi'_a,\mu'\}$ in the five-dimensional minimal supergravity, which is given by
\begin{align}
\begin{split}\label{eq:ctrans-metric}
&\tau' = D^{-1} \tau,\quad
\lambda'_{00} = D^{-2} \lambda_{00},\quad
 \lambda'_{01} =D^{-2} (c^3 \lambda_{01}+s^3 \lambda_{00} \omega_0),\\
& \lambda'_{11} = -\frac{\tau D}{\lambda_{00}} + \frac{(c^3 \lambda_{01}+s^3 \omega_0 \lambda_{00})^2}{D^2 \lambda_{00}},\\
&\omega'_0 = D^{-2}\left[c^3(c^2+s^2+2s^2 \lambda_{00})\omega_0-s^3 (2c^2+(c^2+s^2)\lambda_{00}) \lambda_{01}\right],\\
& \omega'_1 = \omega_1 + D^{-2} s^3 \left[-c^3 \lambda_{01}^2+s(2c^2-\lambda_{00})\lambda_{01} \omega_{0}-c^3 \omega_0^2\right],\\
& \psi'_0 = D^{-1} sc\, (1+\lambda_{00}),\quad
\psi'_1 = D^{-1} sc\, (c\lambda_{01}-s \omega_0),\\
&\mu' = D^{-1}sc\,(c\,\omega_0-s\lambda_{01}),
\end{split}
\end{align}
with
\begin{align}
 D  = c^2+ s^2 \lambda_{00} = 1 + s^2 (1+\lambda_{00}),
\end{align}
where the new parameter $\alpha$ in $(c,s):=(\cosh\alpha,\sinh\alpha)$ is related to the electric charge.
The functions $a_{\phi}^{\prime a}\ (a=0,1)$ and the component $A'_\phi$ for the charged solution are determined by  for the eight scalar functions $\{\lambda'_{ab},\omega'_a,\psi'_a,\mu'\}$ from  Eqs.~(\ref{eq:mu}) and (\ref{eq:twistpotential}) after the replacement of $\{\lambda_{ab},\omega_a,\psi_a,\mu\}$ with $\{\lambda'_{ab},\omega'_a,\psi'_a,\mu'\}$, 
and thus one can obtain the new metric and gauge potential that describe the charged solution for Eqs.~(\ref{Eineq}) and (\ref{Maxeq}).
This transformation adds the electric and dipole charges to a vacuum black ring solution while keeping the asymptotic flatness and Killing isometries.
However, as mentioned in Ref.~\cite{Bouchareb:2007ax}, when one performs the Harrison transformation for the regular vacuum black ring such as the Pomeransky-Sen'kov solution, a Dirac-Misner string singularity inevitably appears on the disk inside the ring, though the transformation can generate the regular Cveti\v{c}-Youm charged black hole for the vacuum black hole such as the Myers-Perry solution.  
In the following, to solve this undesirable problem, we will use the vacuum rotating black ring having a Dirac-Misner string singularity on the disk inside the ring as the seed of the Harrison transformation.

\begin{figure}[t]
\includegraphics[width=7cm]{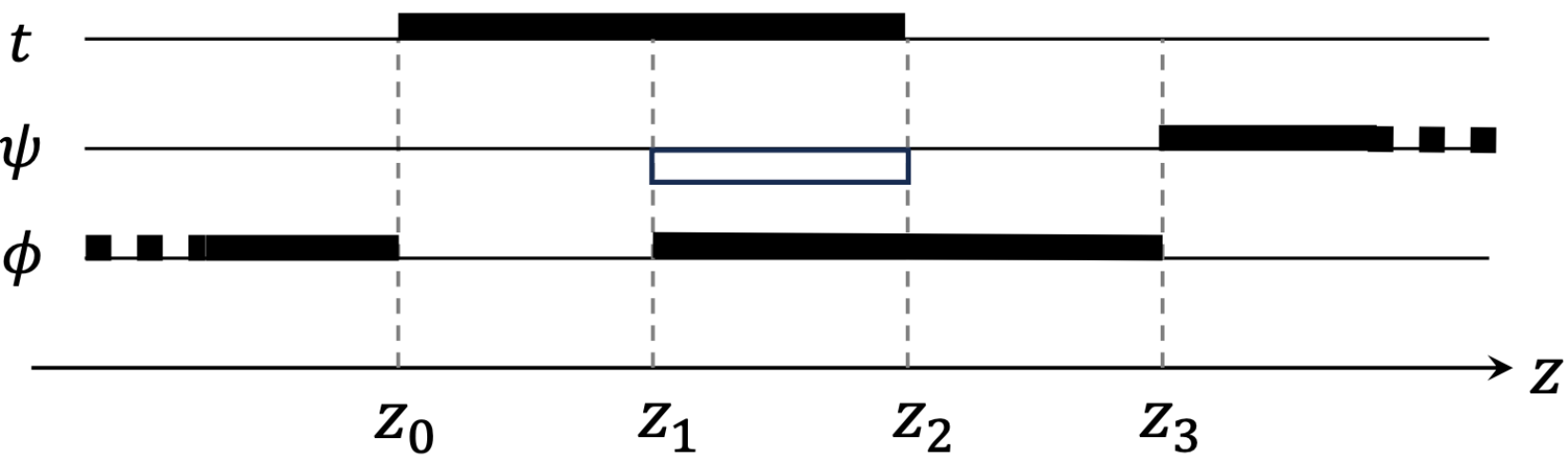}
\caption{Rod structure of the diagonal seed metric. 
The white bar is the negative density rod.\label{fig:rod-seed}}
\end{figure}

\section{ Vacuum seed solution for Harrison transformation}\label{sec:ism}
Since the first work by Pomerasky~\cite{Pomeransky:2005sj}, the ISM~\cite{Belinsky:1979mh,Belinski:2001ph} has been used to construct various vacuum solutions  of five-dimensional black holes~\cite{
Mishima:2005id,
Tomizawa:2005wv,Tomizawa:2006jz,Tomizawa:2006vp,
Iguchi:2006rd,
Elvang:2007rd,
Tomizawa:2007mz,Iguchi:2007xs,
Pomeransky:2006bd,
Iguchi:2007is,Evslin:2007fv,
Elvang:2007hs,Izumi:2007qx,
Chen:2011jb,
Chen:2008fa,
Chen:2012kd,
Rocha:2011vv,
Rocha:2012vs,
Chen:2015iex,
Chen:2012zb,
Lucietti:2020ltw,
Lucietti:2020phh,
Morisawa:2007di,Evslin:2008gx,
Feldman:2012vd,
Chen:2010ih,
Iguchi:2011qi,
Tomizawa:2019acu,
Tomizawa:2022qyd,
Suzuki:2023nqf,
Figueras:2009mc}, with the help of the rod structure~\cite{Harmark:2004rm}.
Here we use the ISM to construct the vacuum seed solution in five dimensions used for the electric Harrison transformation in the following section, i.e.,  a vacuum rotating black ring solution having a Dirac-Misner singularity between the  horizon and the center of the ring.

\medskip
Let us consider an asymptotically flat, stationary and bi-axisymmetric vacuum spacetime with three commuting Killing vector fields, 
a timelike Killing vector field  $\partial/\partial t$ and two spacelike axial Killing vector fields $\partial/\partial \psi$, $\partial/\partial \phi$, then from the integrability of two-planes orthogonal to the Killing vector fields~\cite{Emparan:2001wk,Harmark:2004rm},  
the metric  can be written in the Weyl-Papapetrou form
\begin{align}
ds^2 = G_{ij} dx^i dx^j + f (d\rho^2+dz^2),
\end{align}
with the constraint
\begin{eqnarray}
\det(G_{ij}) = -\rho^2, \label{eq:constraint}
\end{eqnarray}
where $(x^i)=(t,\psi,\phi)$ and the three-dimensional metric $G_{ij}$ and the function $f$ depend on $\rho$ and $z$ only.  
Then, the vacuum Einstein equation reduces to the equation for $3\times 3$ matrix $G=(G_{ij})$
\begin{align}
 \partial_\rho U + \partial_z V = 0,\quad U:=\rho \partial_\rho G G^{-1},\quad V:=\rho \partial_z G G^{-1},\label{eq:eqUV}
\end{align}
and the equation for $f$
\begin{align}
 \partial_\rho \log f = - \fr{\rho} + \fr{4\rho} {\rm tr} (U^2-V^2),\quad \partial_z \log f  = \fr{2\rho} {\rm tr}(UV). \label{eq:f}
\end{align}
Note  that the integrability condition $\partial_{\rho}\partial_zf= \partial_{z}\partial_\rho f$ with respect to $f$ is automatically satisfied for the solution of Eq.~(\ref{eq:eqUV}), and the function $f$ is determined from $G$ up to a constant factor.

\medskip
To our end, let us start with the five-dimensional solution given by the following diagonal metric, whose rod diagram is drawn in Fig.~\ref{fig:rod-seed} under the assumption of $z_0<z_1<z_2<z_3$, 
\begin{align}
G_0 &= {\rm diag} \left( -\frac{\mu_0}{\mu_2},\frac{\mu_2\mu_3}{\mu_1}, \frac{\rho^2 \mu_1}{\mu_0\mu_3}\right),\\
f_0&=C_f\frac{\mu_2 \mu_3 R_{01}R_{02}R_{12} R_{13}^2}{\mu_1  R_{00} R_{03} R_{11} R_{22}R_{23} R_{33}},
\end{align}
where $C_f$ is an arbitrary constant, which is chosen as $1$ throughout this paper,  $\mu_i:=\sqrt{\rho^2+(z-z_i)^2}-z+z_i$ and $R_{ij}:=\rho^2+\mu_i \mu_j$.
Since it is known that the regular black ring can be constructed from a singular diagonal seed with the negative density rod~\cite{Elvang:2007rd,Emparan:2008eg}, we use the similar seed metric that is singular at $\rho=0$ on $z_1<z<z_2$ where a negative density rod is induced\footnote{While the negative density rod is put on the left side of the horizon rod in Refs.~\cite{Elvang:2007rd,Emparan:2008eg},
we induce it on the right side as in Ref~\cite{Chen:2008fa}. In the former case, we have also tried to make the charged black ring using the setup in Ref.~\cite{Chen:2011jb}. But we found that the conditions for having the ring horizon and absence of the Dirac-Misner string singularity cannot hold at the same time except for the vacuum case. }.

\medskip
Following the procedure in Ref.~\cite{Pomeransky:2005sj},
we remove two trivial solitons from the endpoints  $z=z_2,z_3$ with the vectors $(1,0,0)$ and  $(0,0,1)$, respectively, and then obtain the unphysical metric which does not satisfy the constraint~(\ref{eq:constraint})
\begin{eqnarray}
\tilde{G}_0 &=&  G_0\, {\rm diag} \left(-\frac{\mu_2^2}{\rho^2},1,-\frac{\mu_3^2}{\rho^2} \right) \notag\\
&=& {\rm diag} \left(\frac{\mu_0 \mu_2}{\rho^2},\frac{\mu_2\mu_3}{\mu_1}, - \frac{\mu_3 \mu_1}{\mu_0}\right)\notag\\
&=& {\rm diag} \left(-\frac{\mu_0}{\bar{\mu}_2},\frac{\rho^4}{\mu_1 \bar{\mu}_2 \bar{\mu}_3 },\frac{\rho^2 \mu_1}{\mu_0 \bar{\mu}_3}  \right),\label{eq:diagonalseedmod}
\end{eqnarray}
where  $\bar{\mu}_i :=- \sqrt{\rho^2+(z-z_i)^2}-z+z_i$.
Next, we add back two non-trivial solitons to  the endpoints  $z=z_2,z_3$  with the vectors $m_{2,0}=(1,C_2,0)$ and $m_{3,0}=(0,C_3,1)$, respectively,  
and we obtain the two-soliton solution written in the following physical metric satisfying the constraint~(\ref{eq:constraint}),
\begin{align}
G_2 = \tilde{G}_0 - \sum_{i,j=2,3} (\Gamma^{-1})_{ij} \frac{(m_i \tilde{G}_0) \otimes (m_j \tilde{G}_0)}{\mu_i \mu_j}, \label{eq:G2}
\end{align}
where the $2\times2$ matrix $\Gamma_{ij}$ is written as
\begin{align}
 \Gamma_{ij} := \frac{m_i \tilde{G}_0 m_j}{R_{ij}},\quad m_{i} :=m_{i,0} \Psi_0^{-1}(\lambda=\mu_i,\rho,z), \quad (i,j=2,3),\label{eq:defmi}
\end{align}
with the generating matrix made from Eq.~(\ref{eq:diagonalseedmod})
by the replacement $\mu_i \to \mu_i -\lambda, \bar{\mu_i} \to \bar{\mu_i} -\lambda$ and $\rho^2 \to \rho^2 -2\lambda z-\lambda^2$
\begin{align}
\Psi_0(\lambda,\rho,z)= {\rm diag}\left(
- \frac{\mu_0-\lambda}{\bar{\mu} _2-\lambda},
    \frac{ \left(\rho ^2-2 \lambda 
   z-\lambda ^2\right)^2}{\left(\mu_1-\lambda \right) \left( \bar{ \mu} _2-\lambda\right)
   \left(\bar{\mu} _3-\lambda\right)},
  \frac{ \left(\mu _1-\lambda \right) \left(\rho ^2-2 \lambda 
   z-\lambda ^2\right)}{\left(\mu _0-\lambda \right) \left(\bar{\mu} _3-\lambda\right)} 
    \right).
\end{align}
Here, we have introduced $\bar{\mu}_2$ and $\bar{\mu}_3$ in the last line of Eq.~(\ref{eq:diagonalseedmod}) using  $\mu_i \bar{\mu}_i=-\rho^2$, to eliminate $\mu_2$ and $\mu_3$ which cause the divergence in $\Psi_0^{-1}(\mu_2,\rho,z)$ and $\Psi_0^{-1}(\mu_3,\rho,z)$.
The metric function $f$ can be written as
\begin{align}
 f_2 = \frac{{\rm det} (\Gamma_{ij} )}{{\rm det}( \Gamma_{ij})|_{C_2\to 0,C_3\to 0}}f_0. \label{eq:f2}
\end{align}

\medskip

The divergence of the metric on the rod $[z_1,z_2]$ can be removed by setting the BZ parameter $C_2$ as
\begin{align}
C_2 =z_2^2\sqrt{ \frac{2}{z_{20}z_{21}z_{23}}} ,\label{eq:no-negative}
\end{align}
where $z_{ij}:=z_i-z_j$ and this assures that two rod vectors on $[z_1,z_2]$ and $[z_2,z_3]$ becomes parallel so that the point $(\rho,z)=(0,z_2)$ is no longer the endpoint of different rods but becomes a mere regular point which is often referred to as a phantom point.
Hence, the two rods are merged to a single rod $[z_1,z_3]$. 
For later convenience, we also redefine the BZ parameter $C_3$ as
\begin{align}
 a:= \frac{z_{31}^2}{z_3 z_{30}}C_3.
\end{align}
The obtained metric becomes asymptotically flat at infinity 
$\sqrt{\rho^2+z^2}\to\infty$, if and only if 
\begin{align}
 -1 < a < 1,\label{eq:constQ}
\end{align}
and  if not,
the two-dimensional metric $G_{IJ}$ ($I,J=\psi,\phi$) behaves as $G_{IJ}\simeq \tilde G_{IJ}/(1-a^2)$ ($\tilde G_{IJ}$: a certain positive semi-definite metric), and hence the metric does not become Lorenzian, at least,  at infinfty.

\medskip
Under this condition, 
we redefine the coordinates $(x^i)=(t,\psi,\phi)$ so that the metric asymptote to the  Minkowski metric at a rest frame  as
\begin{eqnarray}
x^i \to \Lambda^i{}_j x^j, \label{eq:globalrot}
\end{eqnarray}
\begin{align}
 \Lambda = \arrayL{ccc}  1 & - \Gamma_1 \Gamma_2 &0\\
 0 & \Gamma_1 & -a \Gamma_1  \\
 0 & -a \Gamma_1  & \Gamma_1  \arrayR,\quad \Gamma_1 := \fr{\sqrt{1-a^2}},\quad
 \Gamma_2 := \sqrt{\frac{2z_{20}z_{21}}{z_{32}}}.
\end{align}

Note that the metric after the global rotation has the following rod structure:
\medskip
\begin{itemize}
\item[(i)]{$(-\infty,z_0]$:} a spacelike semi-infinite rod with the rod vector $v_0=(0,0,1)$,

\item[(ii)]{$[z_0,z_1]$:} a timelike finite rod  with the rod vector
\begin{align}
v_{01} = \left(1, \frac{1}{z_{31}-a^2 z_{23}}\sqrt{\frac{z_{32}z_{21}}{2z_{20}}}, \frac{a}{z_{31}-a^2 z_{23}}\sqrt{\frac{z_{32}z_{21}}{2z_{20}}}  \right),
\end{align}

\item[(iii)]{$[z_1,z_3]$:} a finite rod with the rod vector
\begin{align}
v_{13} = \left( a \sqrt{\frac{2z_{20}z_{21}}{z_{32}}},   \frac{a (z_{31}^2-z_{30}z_{32})}{z_{31}^2-a^2 z_{30}z_{32}},1\right).
\end{align}
In order that the horizon cross-section has the topology of $S^2\times S^1$, 
both the $\psi$ and $t$-components  in $v_{13}$ must be zero, 
however, we require for now that the $\psi$-component only vanishes. 
Therefore, we set the phantom point $z=z_2$ as
\begin{align}
 z_2 = z_3 - \frac{z_{31}^2}{z_{30}}.\label{eq:ringcond}
\end{align}
Note that the presence of the remaining $t$-component in the rod vector $v_{13}$ leads to a Dirac-Misner string singularity inside the black ring, which we will eliminate not before but after the Harrison transformation, as preformed in the next section.

\item[(iv)]{$[z_3,\infty)$}: a spacelike semi-infinite rod with the rod vector 
$v_3 = (0,1,0)$\, .
\end{itemize}

\begin{figure}[t]
\includegraphics[width=7cm]{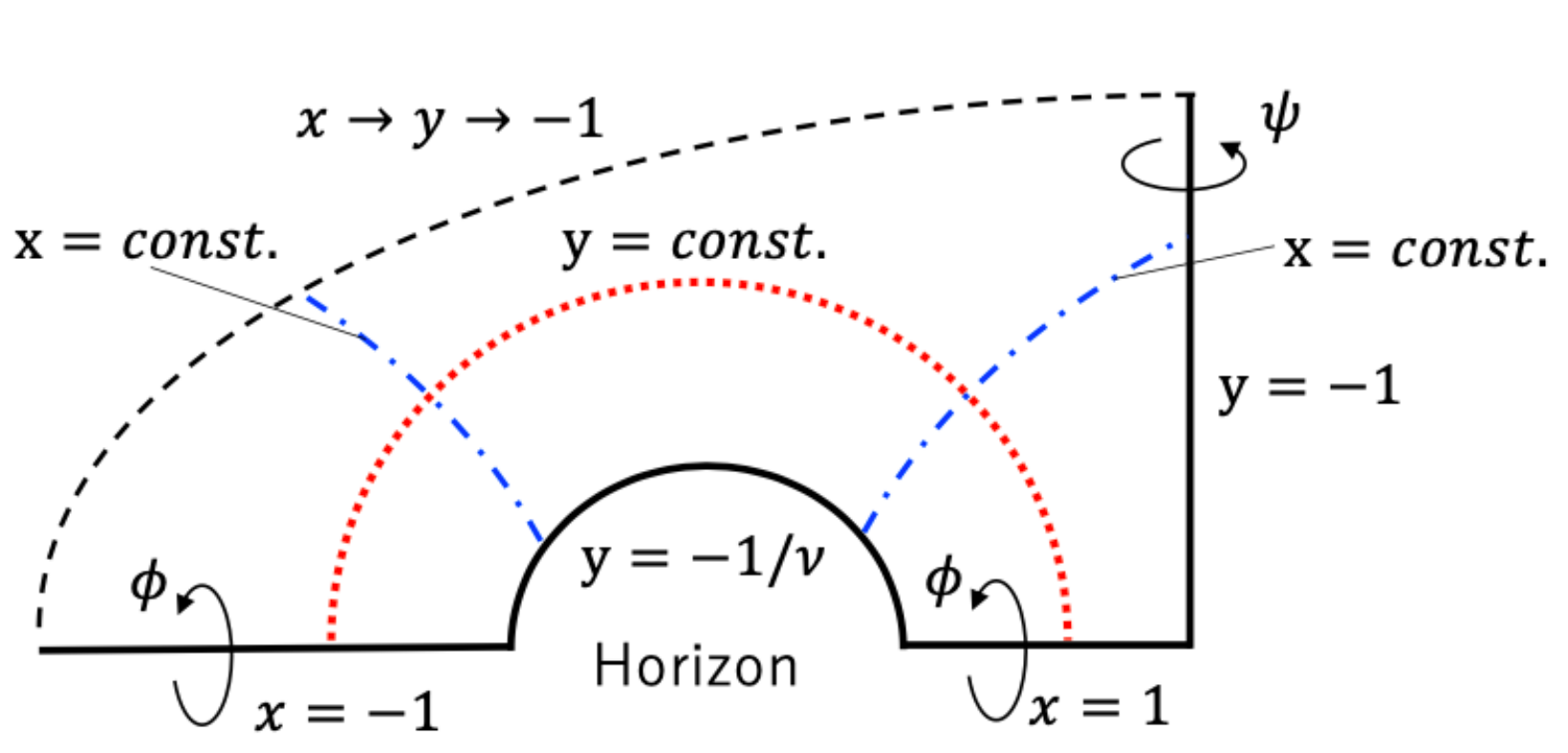}
\caption{
The  $C$-metric coordinates are represented in the orbit space of the black ring. The black dashed curve corresponds to the asymptotic infinity. Additionally, the red dotted and blue dot-dashed curves correspond to surfaces where $y=\text{const.}$ and $x=\text{const.}$, respectively.
\label{fig:cmetric}}
\end{figure}

\subsection{$C$-metric coordinates}

The canonical coordinates $(\rho,z)$ are useful for the solution generation but not suitable  for the analysis  for the black ring, such as the proof of the absence of curvature singularities and CTCs. 
Hence, before performing the Harrison transformation, using Eqs.~(\ref{eq:no-negative}) and (\ref{eq:ringcond}), we transform the canonical coordinates $(\rho,z)$ into  the $C$-metric coordinates $(x,y)$~\cite{Harmark:2004rm} defined by 
\begin{align}
 \rho = \frac{2 \ell^2 \sqrt{-G(x)G(y)}}{(x-y)^2},\quad z = \frac{\ell^2(1-xy)(2+\nu(x+y))}{(x-y)^2},
\end{align}
and instead of the endpoints 
$z_i\ (i=0,1,3)$,  introduce the new parameters $\nu, \ell$ ($0<\nu<1, \ell>0$), 
\begin{align}
 z_0 = - \nu\ell^2,\quad z_1 = \nu \ell^2, \quad z_3 = \ell^2,
 \label{eq:newparamz}
\end{align}
where 
\begin{align}
G(\xi) = (1-\xi^2)(1+\nu \xi).
\end{align}
The functions including nasty square roots, $\mu_0,\mu_1,\mu_3$, are written as rational functions of $x,y$,
\begin{align}
 \mu_0 =- \frac{2\ell^2(1-x)(1+y)(1+\nu y)}{(x-y)^2},\quad
 \mu_1 = - \frac{2\ell^2(1-x)(1+\nu x)(1+y)}{(x-y)^2}\quad
 \mu_3 = \frac{2\ell^2 (1+\nu x)(y^2-1)}{(x-y)^2}.
\end{align}
Then, it can be found from Eq.~(\ref{eq:G2}) and (\ref{eq:f2}) that 
the metric of the vacuum black ring with a Dirac-Misner string singularity can be written as
\begin{align}
&ds^2 = - \frac{H(y,x)}{H(x,y)} (dt+\Omega_\psi(x,y) d\psi + \Omega_\phi(x,y) d\phi)^2+\frac{1}{H(y,x)}\left[F(y,x)d\psi^2-2J(x,y)d\psi d\phi-F(x,y)d\phi^2\right]\nonum
&\quad + \frac{2\ell^2  H(x,y)}{(1-\nu^2)(1-a^2)(x-y)^2}\left(\frac{dx^2}{G(x)}-\frac{dy^2}{G(y)}\right),
\label{eq:metricsol-vac}
\end{align}
where the metric functions are given by
\begin{align}
&H(x,y) = \left(1-\nu ^2\right) \left[\left(1-a^2\right) (1+\nu  x)^2+\nu ^2
   \left(1-x^2\right)\right]+a^2 \nu ^2 \left(y^2-1\right) (\nu  x+1)^2,\label{eq:Hxy}\\
&F(x,y)=-\frac{2\ell^2}{(1-a^2)(x-y)^2}\biggr[\left(1-\nu ^2\right) G(x) \left(\nu ^2+2 \nu  y+1\right)
+a^4 G(x) (1-\nu  y) (1+\nu  y)^3\nonum
&\hspace{3cm}-a^2 \left(\nu ^2 G(y) \left(2 G(x)-\nu ^2
   \left(1-x^2\right)^2\right)+2 \left(1-\nu ^2\right) G(x) (1+\nu   y)^2\right)\biggr],\\
&J(x,y)= \frac{2 a \ell^2 \nu ^3 \left(1-x^2\right) \left(1-y^2\right) 
\left(1-\nu ^2-a^2 (1+\nu  x) (1+\nu    y)\right)}{\left(1-a^2\right) (x-y)},\\
&\Omega_\psi(x,y) = \frac{2\ell \nu(1+\nu) (y+1)}{ \sqrt{(1-\nu^2)(1-a^2)}H(y,x)}\left[a^2 (1+\nu  x) \left(1-\nu ^2+\nu  (1-x) (1+\nu  y)\right)-(1-\nu ) (\nu +1)^2\right],\\
&\Omega_\phi(x,y) = \frac{2a  \ell \nu(1-\nu )  (x+1)}{\sqrt{(1-\nu^2)(1-a^2)} H(y,x)}\left[a^2 (1+\nu  x) (1+\nu  y)^2-(\nu +1) ((1+\nu  x) (1+\nu  y)+(1-\nu ) \nu  (y-1))\right].
\end{align}
We should note that since the point $z=z_2$ on the rod is phantom, as expected, the extra square root term $\sqrt{\rho^2+(z-z_2)^2}$ in $\mu_2$ cancel out due to the regularity condition~(\ref{eq:no-negative})  so that it disappears in the metric. 
Throughout this paper, we assume that the coordinates $(t,\psi,\phi,x,y)$ run the ranges,
\begin{eqnarray}
-\infty < t <\infty, \quad 0\leq \psi\le 2\pi, \quad 0\le \phi \leq 2\pi, \label{eq:tphipsirange}
\end{eqnarray}
and
\begin{eqnarray}
-1\leq x \leq 1, \quad -1/\nu \leq y \leq -1. \label{eq:xyrange}
\end{eqnarray}

\begin{figure}[t]
\includegraphics[width=7cm]{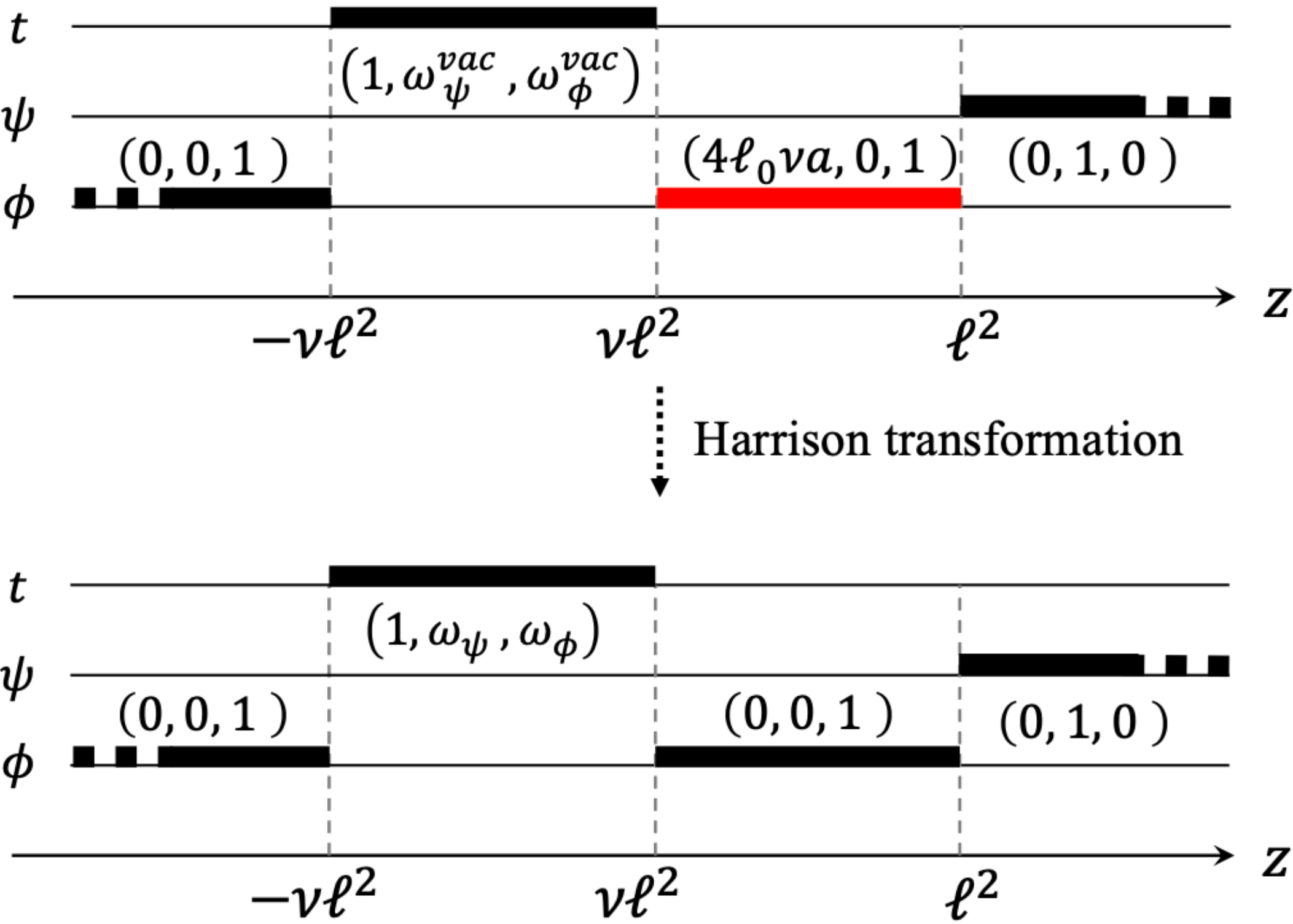}
\caption{
The rod structures before and after the Harrison transformation are depicted. 
The red bar indicates the rod with the Dirac-Misner string singularity. 
The latter is shown after imposing the condition~(\ref{eq:nodms}).
 \label{fig:rod-trans}}
\end{figure}

\medskip
For later convenience, we  give the rod structure for the obtained vacuum black ring with a Dirac-Misner string singularity in the $C$-metric coordinate system, whose  boundaries  can be again described as follows (Fig.~\ref{fig:rod-trans}): 
\begin{itemize}
\item[(i)] $\phi$-rotational axis outside the black ring: $\partial \Sigma_\phi=\{(x,y)|x=-1,-1/\nu<y<-1\}$ with the rod  vector $v_\phi=(0,0,1)$\,,
where in the choice of  $C_f=1$, 
the periodicity $\Delta \phi =2\pi$ of $\phi$ in Eq.~(\ref{eq:tphipsirange})
assures the absence  of  conical singularities on $\partial \Sigma_\phi$,

\item[(ii)] Horizon:  $\partial \Sigma_{\cal H}=\{(x,y)|-1<x<1,y=-1/\nu\}$ with the rod  vector $v_{\cal H}=(1, \omega^{\rm vac}_\psi ,\omega^{\rm vac}_\phi) $, where
\begin{align}
(\omega^{\rm vac}_\psi,\omega^{\rm vac}_\phi) = \frac{\sqrt{\left(1-a^2\right) \left(1-\nu ^2\right)}}{2 \ell   \left(1+\nu - (1-\nu )a^2\right)}(1,a),
\end{align}

\item[(iii)] $\phi$-rotational axis inside the black ring: $\partial \Sigma_{\rm in}=\{(x,y)|x=1,-1/\nu<y<-1\}$ with the rod  vector $v_{\rm in}=(4\ell_0 \nu a,0,1)$ where $\ell_0 := \ell/\sqrt{(1-a^2)(1-\nu^2)}$\,, 

\item[(iv)] $\psi$-rotational axis outside the black ring: $\partial \Sigma_\psi=\{(x,y)|-1<x<1,y=-1\}$ with the rod  vector $v_\psi=(0,1,0)$\,, 
where together with $C_f=1$, the periodicity $\Delta \psi =2\pi$ of $\psi$ in Eq.~(\ref{eq:tphipsirange})
assures the absence  of  conical singularities on $\partial \Sigma_\psi$,

\item[(v)]
Infinity:  
$\partial \Sigma_\infty 
= \{(x,y)|x\to y\to -1 \}$ \, .
 
\end{itemize}
When $a=0$, one reproduces the regular singly-rotating black ring of Emparan and Reall in Ref.~\cite{Emparan:2001wn} . 
When $a\not=0$, there are a Dirac-Misner string singularity~\cite{Misner:1963fr} on $\partial\Sigma_{\rm in}$  due to the presence of the nonzero $t$-component in the rod vector $v_{\rm in}$. Consequently, the time coordinate $t$ must have the period of $8 \pi\ell \nu a$ to avoid the conical singularities on $\partial\Sigma_{\rm in}$.
Hence, one may consider that for $a\not=0$ the solution is unphysical but we do not remove the  Dirac-Misner string singularity at this stage.
In the following section, after applying the electric Harrison transformation~\cite{Bouchareb:2007ax} to
the vacuum rotating black ring having the Dirac-Misner string singularity,
we will control the parameter $a$ to eliminate it.

\section{Charged black ring}\label{sec:sol}
Applying the electric Harrison transformation, Eq.~(\ref{eq:ctrans-metric}) (Eq.~(119) in Ref.~\cite{Bouchareb:2007ax}), to the vacuum solution~(\ref{eq:metricsol-vac}) in Sec.~\ref{sec:ism}, where we choose the two commuting Killing vectors used for the dimensional reduction to three dimensions as $\xi_0:=\partial/\partial t$ and $\xi_1:=\partial/\partial\psi$, 
we obtain the metric and gauge potential for the charged solution as follows:
\begin{align}
&ds^2 = - \frac{H(y,x)}{D^2H(x,y)} (dt+\Omega')^2+\frac{D}{H(y,x)}\left[F(y,x)d\psi^2-2J(x,y)d\psi d\phi-F(x,y)d\phi^2\right]\nonum
&\quad + \frac{2\ell^2 D H(x,y)}{(1-\nu^2)(1-a^2)(x-y)^2}\left(\frac{dx^2}{G(x)}-\frac{dy^2}{G(y)}\right),
\label{eq:metricsol}
\end{align}
and
\begin{align}
&A = \frac{\sqrt{3}cs}{DH(x,y)} \left[ (H(x,y)-H(y,x))dt \right.\nonum
&\hspace{2cm} \left.- (c H(y,x) \Omega_\psi(x,y)-s H(x,y) \Omega_\phi(y,x))d\psi
- (c H(y,x) \Omega_\phi(x,y)-s H(x,y) \Omega_\psi(y,x))d\phi\right],
\end{align}
where $(c,s):=(\cosh\alpha,\sinh\alpha)$ and
\begin{align}
&D := \frac{c^2 H(x,y)-s^2 H(y,x)}{H(x,y)},\\
&\Omega' := \left[c^3 \Omega_\psi(x,y)-s^3 \Omega_\phi(y,x)\right]d\psi + \left[c^3\Omega_\phi(x,y)-s^3\Omega_\psi(y,x)\right]d\phi.
\end{align}

\medskip
Let us see the change of the rod structure caused by the Harrison transformation. 
The boundaries of the $C$-metric coordinates $(x,y)$ for the above charged solution can be described as follows (Fig.~\ref{fig:rod-trans}): 
\begin{itemize}
\item[(i)] $\phi$-rotational axis outside the black ring: $\partial \Sigma_\phi=\{(x,y)|x=-1,-1/\nu<y<-1\}$ with the rod  vector $v_\phi=(0,0,1)$\,, 

\item[(ii)] Horizon:  $\partial \Sigma_{\cal H}=\{(x,y)|-1<x<1,y=-1/\nu\}$ with the rod  vector $v_{\cal H}=(1, \omega_\psi, \omega_\phi)$, where
\begin{align}
(\omega_\psi,\omega_\phi) = \frac{\sqrt{\left(1-a^2\right) \left(1-\nu ^2\right)}}{2 \ell \left(c^3
   \left(1+\nu-  (1-\nu)a^2\right)-2 a \nu  s^3\right)}(1,a),
\end{align}

\item[(iii)]  
Inner axis of the black ring: $\partial \Sigma_{\rm in}=\{(x,y)|x=1,-1/\nu<y<-1\}$ with the rod  vector $v_{\rm in}=( 4\ell_0 \nu( c^3 a -s^3),0,1)$

\item[(iv)] $\psi$-rotational axis outside the black ring: $\partial \Sigma_\psi=\{(x,y)|-1<x<1,y=-1\}$ with the rod  vector $v_\psi=(0,1,0)$\,, 

\item[(v)]
Infinity:  
$\partial \Sigma_\infty 
= \{(x,y)|x\to y\to -1 \}$ \,, 
 
\end{itemize}

It can be seen that the Harrison transformation does not change both the positions of the $\phi$-rotational axis  $\partial\Sigma_\phi$ at $x=-1$ and the $\psi$-rotational axis $\partial\Sigma_\psi$ at  $y=-1$, and also leaves both the directions of the rod vectors invariant because it preserves asymptotic flatness.  Furthermore, it does not affect the regularity, specifically the absence of conical singularities for the coordinate ranges (\ref{eq:tphipsirange}),  on the two axes $\partial \Sigma_\phi$ and $\partial \Sigma_\psi$,
This transformation does not  also change the positions of the horizon $\partial \Sigma_{\cal H}$ at $y=-1/\nu$ and  the inner axis inside the black ring $\partial \Sigma_{\rm in}$ at $x=1$ but changes the directions of the rod vectors. 
$v_{\cal H}$ is changed from $(1,\omega^{\rm vac}_\psi,\omega^{\rm vac}_\phi)$ to $(1,\omega_\psi,\omega_\phi)$ and $v_{\rm in}$  from
 $( 4\ell_0 \nu a ,0,1)$ into $( 4\ell_0 \nu ( c^3 a -s^3),0,1)$,
where  the presence of the $t$-component in $v_{\rm in}$ still yields a Dirac-Misner string singularity.
However, at this point one can remove the Dirac-Misner string singularity by setting 
\begin{align}
   a = \tanh^3\alpha.\label{eq:nodms}
\end{align}
It should be noted that this removal is possible only when the vacuum seed solution (corresponding to $(c,s)=(1,0)$) for the  Harrison transformation has a Dirac-Misner string singularity. This is because
the rod vector $v_{\rm in}$ is changed to $( -4 \ell_0 \nu s^3,0,1)$
 by the transformation, and accordingly, the Dirac-Misner string singularity inevitably appears if we choose the vacuum seed solution that does not have the Dirac-Misner string singularity (corresponding to $a=0$), as in Ref.~\cite{Bouchareb:2007ax}.
After removing the Dirac-Misner string singularity by the condition (47), it can be immediately found  that the periodicity $\Delta \phi=2\pi$ of $\phi$ in Eq.~(\ref{eq:tphipsirange}) automatically assures that
there are  no conical singularities on the inner axis $\partial \Sigma_{\rm in}$ of the black ring,

\medskip
In the following, instead of $\alpha$, we use the new parameter for convenience
\begin{align}
\beta := \tanh \alpha.
\end{align}
Since the solution is invariant under the $Z_2$-symmetry $ \beta \to -\beta, \ a\to -a,\ \phi \to - \phi$, we can also assume $0\leq \beta<1$ without loss of generality. Therefore, the solution is described by the three independent parameters $(\ell,\nu,\beta)$ whose ranges are given by
\begin{align}
\ell >0,\quad 0\leq \beta <1,\quad 0<\nu <1. \label{eq:param-region}
\end{align}

\subsection{Absence of curvature singularities}

Now, we demonstrate that within the parameter range (\ref{eq:param-region}), the charged black ring solution exhibits no curvature singularities both on and outside the horizon. 
If curvature singularities were to exist, they would only appear at points where the metric (\ref{eq:metricsol}) or its inverse diverges. 
This divergence occurs solely on the surfaces $H(x,y)=0$, $D=0$, and at the boundaries of the $C$-metric coordinates $x=\pm 1$, $y=-1$, $y=-1/\nu$.

\medskip

In particular, if either the surface $H(x,y)=0$ or the surface $D=0$ were to exist, it would lead to a curvature singularity since the Kretchmann scalar behaves as $R_{\mu\nu\rho\sigma}R^{\mu\nu\rho\sigma}\propto D^{-6}H^{-6}(x,y)$. However, these surfaces do not exist on or outside the horizon, as one can directly observe $H(x,y)>0$ within the ranges (\ref{eq:xyrange}) from Eq.~(\ref{eq:Hxy}). Additionally, $D>0$ within the ranges (\ref{eq:xyrange}) is immediately evident from $D=1+s^2(H(x,y)-H(y,x))/H(x,y)\geq1$, where
\begin{align}
H(x,y)-H(y,x)=2\nu (x-y)
\left[(1-\nu^2)(1-a^2)+ a^2\nu(1-x) (1+\nu y)+ a^2\nu(1+\nu)(-1-y)\right] \geq 0.
\end{align}
One might consider that $H(y,x)=0$ could also potentially lead to divergence in the metric (\ref{eq:metricsol}) at certain points. However, it turns out that $H^{-1}(y,x)$ does not appear in any components of $g_{\mu\nu}$ or $g^{\mu\nu}$. 
As discussed later, the surface $H(y,x)=0$ corresponds to an ergosurface.

\medskip

Moreover, at each boundary, one can also show the absence of curvature singularities by introducing the appropriate coordinates as follows:

\begin{enumerate}[(a)]

\item 

The limit $x\to y \to -1$ corresponds to asymptotic infinity. 
In terms of the standard spherical coordinates $(r,\theta)$, defined as, 
\begin{align}
 x=-1+4(1-\nu)\ell^2 r^{-2} \cos^2\theta ,\quad
  y=-1-4(1-\nu)\ell^2 r^{-2} \sin^2\theta, \label{eq:aslim}
\end{align}
we can find that the metric at $r\to\infty$ ($x\to y \to -1$) behaves as the Minkowski metric
\begin{align}
ds^2 \simeq -dt^2+dr^2 + r^2 (d\theta^2+\sin^2\theta d\psi^2+\cos^2\theta d\phi^2).\label{eq:aslimds}
\end{align}
Hence, the charged solution describes an asymptotically flat spacetime.

\item 

The point $(x,y)=(1,-1)$ corresponds to the center of the black ring, i.e., the intersecting point of the $\psi$-rotational axis and inner $\phi$-rotational axis.
 Using  the coordinates $(r,\theta)$  introduced by
\begin{align}
x = 1 - \frac{(1+\nu)r^2 \cos^2\theta }{\ell^2 (c^2(1+\nu)^2-(1-\nu)^2 s^2)},\quad
y = -1 - \frac{(1+\nu) r^2 \sin^2\theta }{\ell^2 (c^2(1+\nu)^2-(1-\nu)^2 s^2)},
\end{align}
we can show that the metric at $r\to 0$ ($(x,y)\to (1,-1)$) behaves as the origin of the Minkowski spacetime written in the spherical coordinates,
\begin{align}
ds^2 \simeq -dt'^2+dr^2 + r^2 ( d\theta^2 + \sin^2 \theta d\psi^2 + \cos^2 \theta d\phi^2),
\end{align}
where $t' := (1-\nu^2)t/(c^2 (1+\nu)^2-s^2(1-\nu)^2 )$. 
Therefore, the point $(x,y)=(1,-1)$ is regular.

\item

The boundaries $x=-1$ and $x=1$ correspond to the $\phi$-rotational axes outside and inside the black ring, respectively.
Introducing the radial coordinate $r$ by  $x = \pm 1 \mp C_\pm r^2$ with a positive constant $C_\pm$ for $x=\pm1$,
we can see that the metric at $r\to0$ ($x\to \pm 1$) behaves as
\begin{align}
 ds^2 \simeq \gamma^{\pm}_{tt}(y) dt^2+2 \gamma^{\pm}_{t\psi}(y) dt d\psi+ \gamma^{\pm}_{\psi\psi}(y) d\psi^2 + \alpha_\pm(y) ( dr^2+r^2 d\phi^2-G^{-1}(y) dy^2),
\end{align}
where 
\begin{align}
&\gamma_{tt}^{\pm} =-\frac{D|_{x=\pm1}H(y,\pm1)}{H(\pm1,y)},\quad
\gamma_{t\psi}^{\pm} = -\frac{H(y,\pm1)[c^3 \Omega_\phi(\pm1,y)-s^3 \Omega_\phi(y,\pm1)]}{D^2|_{x=\pm1}H(\pm1,y)},\nonum
&\gamma_{\psi\psi}^{\pm} =\frac{D|_{x=\pm1} F(y,\pm1)}{H(y,\pm1)}-\frac{H(y,\pm1)[c^3\Omega_\psi(\pm1,y)-s^3 \Omega_\phi(y,\pm1)]^2}{D^2|_{x=\pm} H(\pm1,y)},\quad
 \alpha_{\pm} = \frac{4 C_\pm \ell^2 D |_{x=\pm 1} H(\pm 1,y)}{(1\pm \nu)(1-\nu^2)(1-a^2)(1\mp y)^2}.
\end{align}
One can also show
\begin{align}
{\rm det} \, (\gamma^{\pm})
 =- \frac{2\ell^2 (1 \pm \nu)^3(1\mp \nu)(1-a^2)(y \pm 1)(1+\nu y)}{(y \mp 1)D |_{x=\pm 1} H(\pm 1,y)}<0.
\end{align}
From $H(x,y)>0$ and $D>0$ for the ranges~(\ref{eq:xyrange}), it is obvious that $\alpha_\pm$ is a positive definite function and $\gamma^{\pm}$ is a nonsingular and non-degenerate matrix for $-1/\nu<y<-1$. Therefore, the metric is regular at $x=\pm1$.

\item 

The boundary $y=-1$ corresponds to the $\psi$-rotational axis. 
Introducing the radial coordinate $r$ by $y=-1-C_0 r^2$ with a positive constant $C_0$, 
we can see that the metric at $r \to 0$ ($y\to -1$) behaves as
\begin{align}
 ds^2 \simeq \gamma^0_{tt}(x)dt^2 + 2 \gamma^0_{t\phi}(x) dt d\phi + \gamma^0_{\phi\phi}(x)d\phi^2 + \alpha_0(x)
 (dr^2+r^2 d\psi^2+G^{-1}(x)dx^2),
\end{align}
where 
\begin{align}
&\gamma_{tt}^{0} =-\frac{D|_{y=-1}H(-1,x)}{H(x,-1)},\quad
\gamma_{t\phi}^{0} = -\frac{H(-1,x)[c^3 \Omega_\phi(x,-1)-s^3 \Omega_\psi(-1,x)]}{D^2|_{y=-1}H(x,-1)},\nonum
&\gamma_{\phi\phi}^{0} =-\frac{D|_{y=-1} F(x,-1)}{H(-1,x)}-\frac{H(-1,x)[c^3\Omega_\phi(x,-1)-s^3 \Omega_\psi(-1,x)]^2}{D^2|_{y=-1} H(x,-1)},\quad
 \alpha_0 = \frac{4 C_0 \ell^2 D|_{y=-1} H(x,-1)}{(1-\nu)^2(1+\nu)(1-a^2)(x+1)^2}.
\end{align}
One can also show
\begin{align}
{\rm det}\,(\gamma^0)
=- \frac{2\ell^2 (1 - \nu)^3(1 + \nu)(1-a^2)(1-x)(1+\nu x)}{(1+x) D |_{y=-1} H(x, -1)}<0.
\end{align}
From $H(x,y)>0$ and $D>0$ for the ranges~(\ref{eq:xyrange}), it is obvious that $\alpha_0$ is a positive definite function and $\gamma^{0}$ is a nonsingular and non-degenerate matrix for $-1<x<1$. Therefore, the metric is also regular at $y=-1$.

\item 

The boundary $y=-1/\nu$ corresponds to the event horizon with the surface gravity
\begin{align}
\kappa = \frac{(1-\nu)(1-\beta ^2)^{3/2} \sqrt{1-\beta ^6}}{4\ell \nu},\label{eq:kappa}
\end{align}
and the null generator is given by $v_{\cal H} = \partial/\partial t + \omega_\psi \partial/\partial\psi + \omega_\phi \partial/\partial \phi$ with 
\begin{align}
(\omega_\psi,\omega_\phi) = \frac{(1-\beta ^2)\sqrt{1-\nu}}{2\ell \sqrt{(1+\nu)(1+\beta ^2+\beta ^4)}}(1,\beta ^3)\label{eq:omegai}.
\end{align}
One can show that $y=-1/\nu$ is a regular Killing horizon by introducing the ingoing/outgoing Eddington-Finkelstein coordinates by
\begin{align}
 dx^i = dx'^i \pm v_{\cal H}^i \frac{(1-\nu^2)}{2 \nu \kappa G(y)} dy, 
\end{align}
where $x^i=(t,\psi,\phi)$ and the metric near $y=-1/\nu$ behaves as
\begin{align}
&ds^2 \simeq \alpha_H(x) \left(\frac{4 \nu^2 \kappa^2 G(y) }{(1-\nu^2)^2}dt'^2 \pm \frac{4\nu \kappa}{1-\nu^2} dt' dy+\frac{dx^2}{G(x)} \right)\nonum
&+ \gamma^H_{\psi\psi}(x)(d\psi'-\omega_\psi dt')^2+ 2\gamma^H_{\psi\phi}(x)(d\psi'-\omega_\psi dt')(d\phi'-\omega_\phi dt')
+ \gamma^H_{\phi\phi}(x)(d\phi'-\omega_\phi dt')^2,
\end{align}
with
\begin{align}
\begin{split}
&\gamma_{\psi\psi}^{H} =\frac{D|_{y=-1/\nu} F(-1/\nu,x)}{H(-1/\nu,x)}
-\frac{H(-1/\nu,x)[c^3\Omega_\psi(x,-1/\nu)-s^3 \Omega_\phi(-1/\nu,x)]^2}{D^2|_{y=-1/\nu} H(x,-1/\nu)},\\
&\gamma_{\psi\phi}^{H} =-\frac{D|_{y=-1/\nu} J(x,-1/\nu)}{H(-1/\nu,x)}-\frac{H(-1/\nu,x)[c^3\Omega_\psi(x,-1/\nu)-s^3 \Omega_\phi(-1/\nu,x)][c^3\Omega_\phi(x,-1/\nu)-s^3 \Omega_\psi(-1/\nu,x)]}{D^2|_{y=-1/\nu} H(x,-1/\nu)},\\
&\gamma_{\phi\phi}^{H} =-\frac{D|_{y=-1/\nu} F(x,-1/\nu)}{H(-1/\nu,x)}-\frac{H(-1/\nu,x)[c^3\Omega_\phi(x,-1/\nu)-s^3 \Omega_\psi(-1/\nu,x)]^2}{D^2|_{y=-1/\nu} H(x,-1/\nu)},\\
& \alpha_H =\frac{2 \nu^2 \ell^2 D|_{y=-1/\nu}  H\left(x,-1/\nu \right)}{\left(1-a^2\right) (1-\nu^2) \left(1+\nu x\right)^2},
\end{split}
\end{align}
and hence, from these we can show
\begin{align}
{\rm det}\, (\gamma^H)
= \frac{8c^6 \nu^2 \ell^4 (1 + \nu)^3(1 - \nu)(1-x^2)}{(1+\nu x) D|_{y=-1/\nu} H(x, -1/\nu)}>0.
\end{align}
It can be seen from $H(x,y)>0$ and $D>0$ for the ranges~(\ref{eq:xyrange}) that $\alpha_H$ is a positive definite function and $\gamma^{H}$ is a nonsingular and non-degenerate matrix for $-1<x<1$.
Hence, the metric is smoothly continued to $-\infty<y<-1/\nu$ across the horizon $y=-1/\nu$. 
Moreover, in the Eddington-Finkelstein coordinate, the gauge potential also remains regular at the horizon $y=-1/\nu$ under the gauge transformation
\begin{align}
 A' =  A \pm d\left( \frac{(1-\nu^2)\Phi_e}{2\nu \kappa} \int \frac{dy}{G(y)}\right),
\end{align}
where $\Phi_e$ is the electric potential defined by
\begin{align}
\Phi_e :=  - (A_t+A_\psi \omega_\psi+A_\phi \omega_\phi)\biggr|_{y=-1} = - \frac{\sqrt{3} \,\beta   \left(1+\beta  ^2+\beta  ^4+\nu  \left(1+\beta  ^2-\beta  ^4\right)\right)}{(\nu +1) \left(1+\beta  ^2+\beta  ^4\right)}.
\label{eq:phi-e}
\end{align}

\item Near the inner and outer rims of the ring horizon at $(x,y)=(\pm1,-1/\nu)$, the spacetime is locally described by the Rindler spacetime. 
By introducing the coordinates $(r,\theta)$ 
\begin{align}
  x = \pm 1 \mp \frac{(1\pm\nu)\kappa R_{1,\mp}}{2(1\mp\nu)\nu \ell^2} r^2 \sin^2\theta,\quad y =-\fr{\nu} \left(1- \frac{(1\pm\nu)\kappa R_{1,\mp}}{4\nu \ell^2} r^2 \cos^2\theta\right),
\end{align}
the metric at $r=0$ ($(x,y)=(\pm1,-1/\nu)$) behaves as
\begin{align}
 ds^2 \simeq dr^2 + r^2 d\theta^2 + r^2 \sin^2\theta (d\phi-\omega_\phi dt)^2  - r^2 \kappa^2 \cos^2 \theta dt^2 + R_{1,\mp}^2 (d\psi-\omega_\psi dt)^2,
\end{align}
where $R_{1,\pm}$ is the $S^1$-radii of the outer and inner rims given by Eq.~(\ref{eq:def-R1pm}), respectively.
In the Cartesian coordinates $(T, X, Y, Z,W)=(\kappa t, r\cos\theta, r\sin\theta \cos(\phi-\omega_\phi t), r\sin\theta \sin(\phi-\omega_\phi t),R_{1,\mp} (\psi-\omega_\psi t) )$, the above asymptotic metric becomes
\begin{align}
 ds^2 \simeq - X^2 dT^2 + dX^2 + dY^2 + dZ^2 +dW^2,
 \end{align}
 where the Rindler horizon lines at $X=0$.
Therefore, the metric is regular at $(x,y)=(\pm1,-1/\nu)$.

\end{enumerate}

\subsection{Absence of closed timelike curves}

Closed timelike curves are absent if the two-dimensional metric $g_{IJ} (I,J=\psi,\phi)$ is positive definite except
for the axes at $x=\pm1$ and $y=-1$.  This is equivalent to the condition ${\rm det}(g_{IJ})>0$ and ${\rm tr}(g_{IJ})> 0$ there.
However, in the current setup, 
${\rm det}(g_{IJ})$ and ${\rm tr} (g_{IJ})$ are clearly positive at the asymptotic infinity at $(x,y)=(-1,-1)$ and also continuous on and outside the horizon. Hence, it suffices to show ${\rm det}(g_{IJ})>0$,
since if there is a point $(x_0,y_0) \in (-1,1) \times [-1/\nu,-1)$ where ${\rm tr}(g_{IJ})\leq 0$, 
then there must be a point $(x_1,y_1)\in (-1,1) \times [-1/\nu,-1)$ where ${\rm tr}(g_{IJ}) =0$ on a curve $\gamma \subset (-1,1)\times [-1/\nu,-1)$ that connects $(x_0,y_0)$ to the asymptotic infinity,
at which two eigenvalues of $g_{IJ}$, $\lambda_1,\lambda_2$,  have opposite signs or both become zero. But either case contradicts the assumption ${\rm det}(g_{IJ})=\lambda_1\lambda_2>0$.

\medskip

To demonstrate the positive definiteness of $\mathrm{det}(g_{IJ})$ away from the axes, it is convenient to remove the zeros at $x=\pm 1$ and $y=-1$. Additionally, addressing the divergent behavior at $x\to y\to -1$, we instead consider the positive definiteness of $\Delta(x,y)$ for $-1 \leq x \leq 1$ and $-1/\nu \leq y \leq -1$, where

\begin{align}
 \frac{4\ell^2(1-x^2)(-1-y)}{(1-\nu^2)(x-y)^4 D H(x,y)} \Delta(x,y):= {\rm det}(g_{IJ}).
\end{align}

\begin{figure}[t]
\includegraphics[width=5cm]{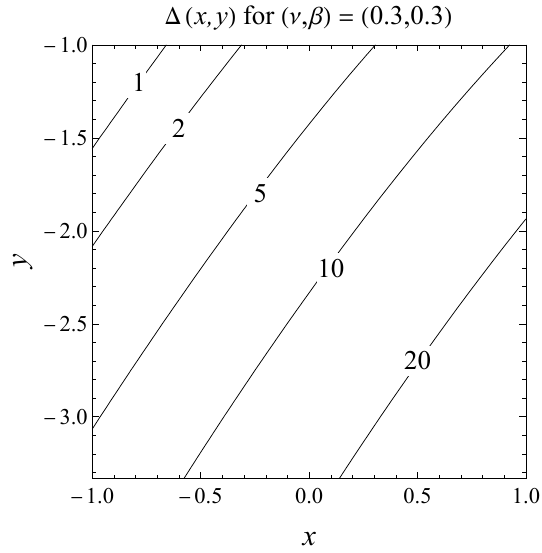}
\caption{
The profile of $\Delta(x,y)$ for $(\nu,\beta)=(0.3,0.3)$ in the C-metric coordinates $(x,y)$. 
Other parameter choices yield similar profiles.
\label{fig:noctc}}
\end{figure}

It is difficult to prove the positivity of $\Delta(x,y)$ in the entire region, but one can easily show the positivity on the horizon
\begin{align}
\Delta(x,-1/\nu) = 2 \nu^{-1}(1-\nu)(1+\nu)^4(1-\beta^2)^{-3}(1+\nu x)^3 >0.
\end{align}

The positivity at $(x,y)=(\pm 1,-1)$ is also evident from the fact that the metric approaches the flat Minkowski metric there. For other regions, we have numerically verified $\Delta(x,y)>0$ both on and outside the horizon for several values of $(\nu,\beta)$ within the parameter range specified in Eq.~(\ref{eq:param-region}) (See Fig. \ref{fig:noctc}).

\section{Physical properties of the charged dipole black ring}\label{sec:phase}

Now, we study the physical properties, the ergoregion, the phase diagram and the shape of the horizon, of the obtained charged dipole black ring obtained in the previous section.

\subsection{Ergoregion}\label{sec:ergo}

The ergosurface of the charged dipole black ring corresponds to the timelike surface $H(y,x)=0$, which are simply solved as
\begin{align}
 y=y_{\pm}(x) := \frac{1-\beta^6 -\nu^2(1-\beta^6 x^2) \pm (1-\nu^2) \sqrt{1-\beta^6(1-\nu^2 x^2)}}{\nu \beta^6(1-\nu^2x^2)}. 
\end{align}
The branch $y=y_+(x)$ does not constitute the ergosurface due to  $y_+(x)>-1$ (refer to the range (\ref{eq:xyrange})), and therefore we consider the other branch $y=y_-(x)$, where  $H(y,x)>0$ for $y_-(x)<y\leq -1$ and $H(y,x)<0$ for $-1/\nu\leq y<y_-(x)$. 
Consequently, this charged dipole black ring admits two types of the ergoregion, depending on the parameters $\nu,\beta$:

\begin{enumerate}[(i)]
\item $-1/\nu<y_-(x)<-1$ for $-1\leq x \leq 1$  when $\nu^2+\beta^6<1$, 
\item $-1/\nu<y_-(x)\leq -1$ for $\sqrt{\nu^{-2}+\beta^{-6}-1}\leq |x|\leq 1$ and $y_-(x)>-1$ for $|x|<\sqrt{\nu^{-2}+\beta^{-6}-1}$ when $\nu^2+\beta^6>1$.
\end{enumerate}

\begin{figure}
\includegraphics[width=5.5cm]{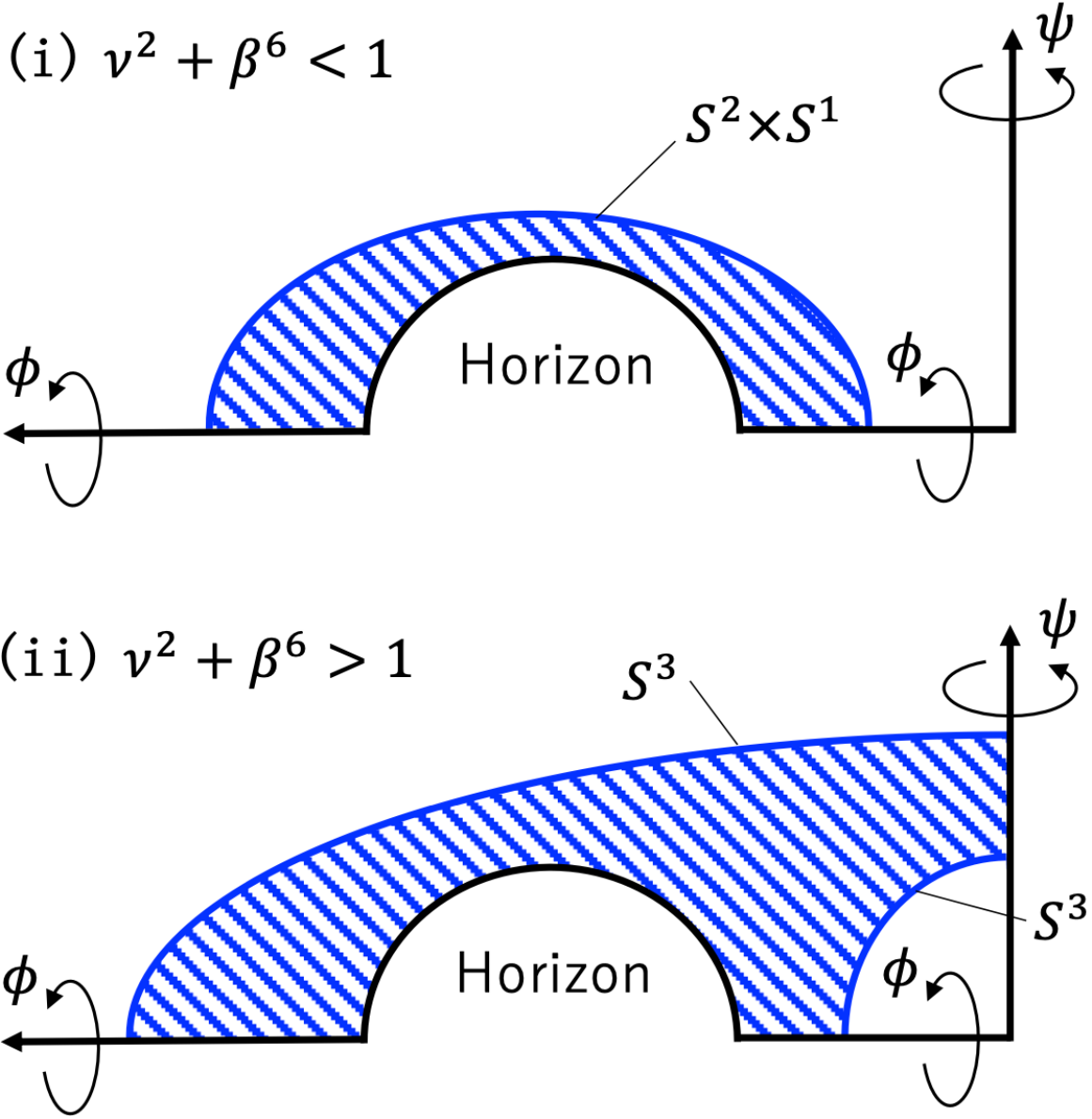}
\caption{
The shapes of  the ergoregion. The ergoregion of charged black rings is depicted with the blue shaded region in the orbit space of the black ring  in case~(i) (small charge $\beta$ with fixed $\nu$) and case~(ii) (sufficiently large charge $\beta$ with fixed $\nu$).
\label{fig:ergo}}
\end{figure}

As depicted in Fig.~\ref{fig:ergo}, in case~(i) (small charge $\beta$ with fixed $\nu$), the ergosurface, with the topology of $S^2\times S^1$, surrounds the ring horizon and intersects the $\phi$-rotational axis but does not touch the horizon or the $\psi$-rotational axis. In case~(ii) (sufficiently large charge $\beta$ with fixed $\nu$), two ergosurfaces are present. The outer ergosurface, with the topology of $S^3$, intersects both the $\phi$ and $\psi$-rotational axes and surrounds both the horizon and the ring center. Conversely, the inner ergosurface, also with the topology of $S^3$, intersects both the $\phi$ and $\psi$-rotational axes but surrounds only the ring center. Since the ring center acts as the fixed point for rotations around the $\psi$ and $\phi$-rotational axes, the ergoregion cannot appear around this point. This is due to the horizon having a larger spin in the $\phi$-direction (\ref{eq:omegai}) as the electric charge increases. The same topology transition of the ergoregion is observed in the vacuum black ring with $S^2$-rotation \cite{Durkee:2008an}.

\subsection{Phase diagram}

Next, let us study the thermodynamic phase of the charged dipole black ring.
We can read off the asymptotic charges from the asymptotic behavior at  $r\to \infty$ with Eq.~(\ref{eq:aslim}):
\begin{align}
 ds^2 =-\left(1 - \frac{8 G_5 M}{3\pi r^2}\right)dt^2- \frac{8 G_5 J_\psi \sin^2\theta}{\pi r^2} dt d\psi
 - \frac{8G_5 J_\phi \cos^2\theta}{\pi r^2} dt d\phi + dr^2 + r^2 ( d\theta^2 + \sin^2\theta d\psi^2+\cos^2\theta d\phi^2).
\end{align}
where  the ADM mass and two ADM angular momenta are written as
\begin{align}
 &M = \frac{3\pi  \nu \ell^2 (1+\nu-(1-\nu)\beta ^6)}{G_5(1-\nu)(1-\beta ^6)} \frac{1+\beta ^2}{1-\beta ^2},\\
 & J_\psi = \frac{2 \pi  \nu \ell^3  \left(1+\nu-(1-\nu) \beta ^6\right) \left((1+\nu)^2-(1-\nu)^2 \beta ^6\right)}{G_5(1-\nu)^{3/2}(1-\beta ^6)^{3/2}(1-\beta ^2)^{3/2}} ,\\
 & J_\phi = \frac{4 \pi  \nu^2 \ell^3 \beta ^3\left((3-\nu) (1+\nu)-(1-\nu) (\nu+3) \beta ^6\right)}{G_5(1-\nu)^{3/2}(1-\beta ^6)^{3/2}(1-\beta ^2)^{3/2}}.
\end{align}
The electric charge evaluated on an arbitrary closed three-surface $S$ enclosing the black ring horizon and the $\phi$-rotational axis $\partial\Sigma_{\rm in}$, identical to that evaluated at spatial infinity $S_\infty$, can be expressed as:
\begin{eqnarray}
Q &:=& \frac{1}{8\pi G_5} \int_{S} \left(\star F + \fr{\sqrt{3}} F \wedge A\right)\notag\\
    &=&  \frac{1}{8\pi G_5} \int_{S_\infty} \star F \notag\\
   &=& - \frac{4M}{\sqrt{3}} \frac{\beta }{1+\beta ^2}\notag\\
   &=& - \frac{2M\tanh(2\alpha)}{\sqrt{3}},\label{eq:defQ}
\end{eqnarray}
where, in the second equality, we have used the fact that the Chern-Simons term diminishes more rapidly than the first term as $r \to \infty$. 
It is evident from the final equation that the mass and electric charge clearly satisfy the BPS bound $M \geq (\sqrt{3}/2)|Q|$, which reaches saturation as $\alpha \to \infty$ ($\beta \to 1$). 
It is important to note that the charge $Q_{\rm H}$ evaluated on the horizon cross-section does not coincide with $Q$ evaluated at spatial infinity $S_{\infty}$\footnote{One can check the same is true for the charged black ring in Ref.~\cite{Elvang:2004xi}.} . This is because $Q_{\rm H}$ and $Q$ are related by:

\begin{align}
Q_{\rm H}:= \fr{8\pi G_5} \int_H (\star F + \fr{{3}}F\wedge A)
=\frac{\beta ^6 (1-\nu )^2+2 \left(1-\beta ^2\right) \beta ^6 (1-\nu ) \nu -(\nu +1)^2}{\left(1+\nu-(1-\nu)\beta ^6\right)
 \left(1+\nu-(1-\nu)\beta ^2  \right)}Q.
\end{align}
The difference  between these two charges arises from the charge evaluated on the three-surface $D_{\epsilon}$ at $x=1-\epsilon$ encompassing the inner $\phi$-rotational axis $\partial\Sigma_{\rm in}$, defined as:
\begin{align}
Q_ {D} :=\fr{8\pi G_5} \int_{D_\epsilon} (\star F + \fr{{3}}F\wedge A)\biggr|_{\epsilon \to 0} = \frac{\beta ^2 \left(1-\beta ^6\right) (1-\nu^2 )}{\left(1+\nu-(1-\nu)\beta ^6\right)
 \left(1+\nu-(1-\nu)\beta ^2  \right)}Q.
\end{align}
It is straightforward to verify that $Q_{\rm H}+Q_{\rm D} = Q$. The black ring also carries the so-called dipole charge defined on the $S^2$ cross-section of the ring,
\begin{align}
q := \fr{4\pi}\int_{S^2} F =- \frac{2\sqrt{3} \nu \ell \beta ^2}{\sqrt{(1-\nu^2)(1-\beta ^2)(1-\beta ^6)}}, 
\end{align}
which is not a conserved charge but characterizes the black ring together with the conserved charges $M,J_\psi,J_\phi,Q$.
The area of the horizon cross-section is computed as
\begin{align}
A_H = \frac{32\pi^2 \ell^3\nu^2}{(1-\nu)(1-\beta ^2)^{3/2}\sqrt{1-\beta ^6}}.
\end{align}

\medskip
As  discusses in Ref.~\cite{Copsey:2005se}, the charged dipole black  ring in five-dimensional minimal supergravity satisfies the Smarr formula and the first law with respect to the physical quantities, $(M,J_\psi,J_\phi,Q,q)$
\begin{align}
  M = \frac{3}{16\pi} \kappa  A_H  +\frac{3}{2} \omega_\psi J_\psi +\frac{3}{2}\omega_\phi J_\phi
  + \frac{1}{2} \Phi_e Q+\fr{2}\Phi_m q, \label{eq:smarr}
\end{align}
and
\begin{align}
 \delta M = \frac{1}{8\pi} \kappa \delta A_H  + \omega_\psi \delta J_\psi +\omega_\phi \delta J_\phi+ \frac{1}{2} \Phi_e \delta Q+\Phi_m \delta q, \label{eq:1stlaw}
\end{align}
where $\Phi_e$ is the electric potential defined in Eq.~(\ref{eq:phi-e})
 and $\Phi_m$ is the `magnetic dipole potential' defined in Refs.~\cite{Copsey:2005se,Kunduri:2013vka}. 
Although the dipole charge is considered an independent parameter in the first law of thermodynamics, it is not a conserved charge obtained from the surface integral at infinity, as initially discussed for the dipole black ring~\cite{Emparan:2004wy}. 
Consequently, since there are an infinite number of black rings specified by a continuous dipole charge,  with the same asymptotic conserved charges, the existence of an independent dipole charge results in the infinite non-uniqueness of the charged dipole black ring. This results in the black ring having much thicker hair compared to black holes in the same theory, which are uniquely specified only by asymptotic conserved charges ~\cite{Tomizawa:2009ua}.
The black ring obtained carries the mass, two angular momenta, an electric charge, and a dipole charge, with only three of these quantities being independent. As a result, the dipole charge can be expressed in terms of other conserved charges. Therefore, it possesses no additional properties (hair) beyond the conserved charges, whereas in contrast, the most general black ring, if it exists, is expected to have an independent dipole charge~\cite{Emparan:2004wy,Feldman:2014wxa}.
 Since the definition of $\Phi_m$ is complicated, instead, we use the Smarr formula~(\ref{eq:smarr}) to evaluate $\Phi_m$.
 Together with Eqs.~(\ref{eq:kappa}) and (\ref{eq:omegai}), the Smarr formula~(\ref{eq:smarr}) leads to
 \begin{align}
  \Phi_m = 0.\label{eq:dipolepot}
 \end{align}
Therefore, the dipole charge does not involve the first law for the obtained solution 
\begin{align}
 \delta M = \frac{1}{8\pi} \kappa \delta A_H  + \omega_\psi \delta J_\psi +\omega_\phi \delta J_\phi+ \frac{1}{2} \Phi_e \delta Q.
\end{align}
One can check the above first law by differentiating with the independent parameters $(\ell,\nu,\beta)$.

\medskip

To  compare the obtained charged dipole black ring with the previously known charged dipole black ring in the following section,  
it is convenient to introduce the dimensionless variables normalized by the mass scale $r_M$ ($M=3\pi r_M^2/(8 G_5)$):
\begin{align}
&j_\psi := \frac{4G_5}{\pi r_M^3}J_\psi = \frac{(1+\nu )^2-(1-\nu )^2 \beta ^6}{2 \sqrt{2} \left(\beta ^2+1\right)^{3/2} \sqrt{\nu  \left(1+\nu-(1-\nu)\beta ^6\right)}},\\
&j_\phi := \frac{4G_5}{\pi r_M^3}J_\phi =\frac{\sqrt{\nu} \beta ^3 \left((3-\nu)(1+\nu)-(1-\nu)(3+\nu)\beta ^6\right)}{\sqrt{2} \ \left(\beta ^2+1\right)^{3/2} \left(1+\nu -(1-\nu) \beta ^6\right)^{3/2}},\\
& \bar{q} := \frac{q}{r_M} = - \frac{\sqrt{2\nu} \beta ^2}{(1+\beta ^2)\sqrt{3(1-\beta ^2+\beta ^4)(1+\nu-(1-\nu)\beta ^6)}},\\
&a_H:=\frac{\sqrt{2}}{\pi^2 r_M^3}A_H =2 \left(1-\beta ^6\right) \sqrt{\frac{(1-\nu ) \nu  (\nu +1)^3}{\left(\beta ^2+1\right)^3 \left(1+\nu -(1-\nu ) \beta ^6\right)^3}}.
\end{align}

\begin{figure}
\includegraphics[width=5.8cm]{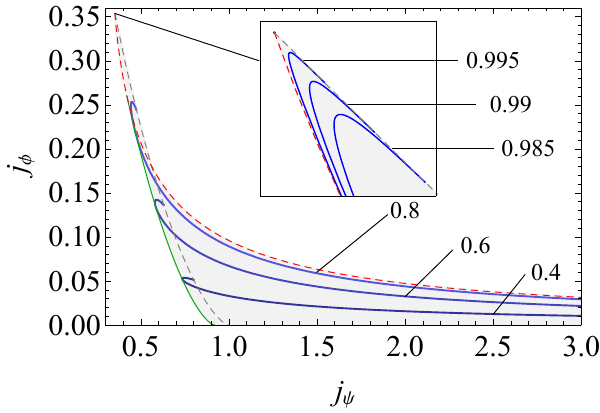}
\includegraphics[width=5.8cm]{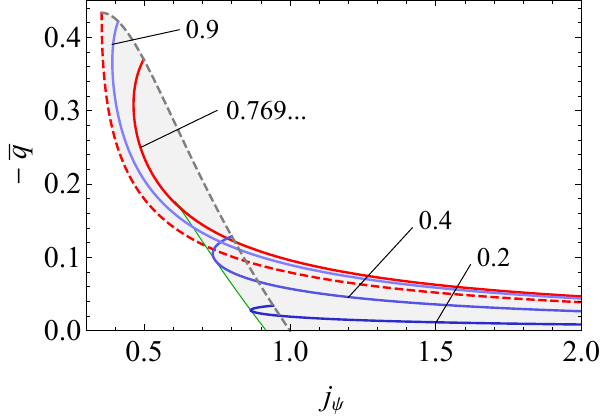}
\includegraphics[width=5.8cm]{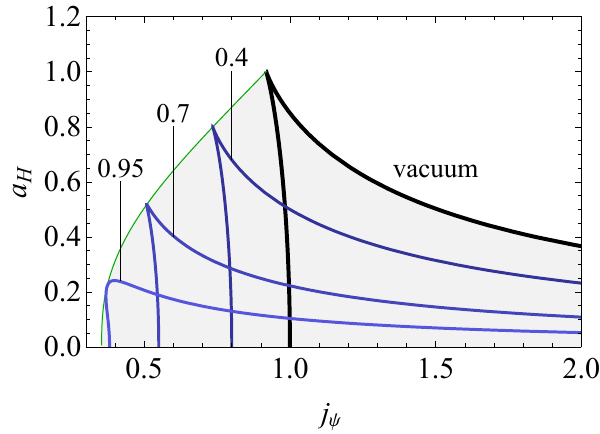}
\caption{
Phase diagram in $(j_\psi,j_\phi)$, $(j_\psi,-\bar{q})$ and $(j_\psi,a_H)$ planes. 
 The colored regions represent the allowed regions.
Each parameter can vary within the following ranges: $1/(2\sqrt{2}) < j_{\psi}$, $0 \leq j_{\phi} < 1/(2\sqrt{2})$, $0 \leq -\overline{q} < (\sqrt{3}/2)$, and $0 < a_H \leq 1$.
The $\beta=\text{const.}$ curves are depicted as blue curves labeled with their respective $\beta$ values. 
Each curve originates from the $\nu=1$ curve (shown as a gray dashed curve in the left two panels and the $j_{\psi}$-axis in the right panel) and extends to the thin ring limit $(j_{\psi}, j_{\phi}, q, a_H) \to (\infty, 0, 0, 0)$ as $\nu \to 0$.
As $\beta \to 1$, the curve asymptotes to the red dashed curves in the left two panels and $a_H=0$ in the right panel. The green curve represents the envelope curve for each $\beta=\text{const.}$ curve.
The green curves  define the boundary of the allowed region. In the middle panel, the red curve corresponds to the branch with $\beta = 0.769...$, establishing the upper bound of $-\overline{q}$ for $j_{\psi}>0.810...$. 
The black curve in the right panel denotes the vacuum black ring phase. \label{fig:phase-newbr}
 }
\end{figure}

In Fig.~\ref{fig:phase-newbr}, we illustrate the allowed regions and some $Q/M={\rm const.}$ phases in the $(j_\psi,j_\phi)$, $(j_\psi,-\bar{q})$ and $(j_\psi,a_H)$ planes.

\subsection{Shape of Ring horizon}

\begin{figure}[t]
\includegraphics[width=5.5cm]{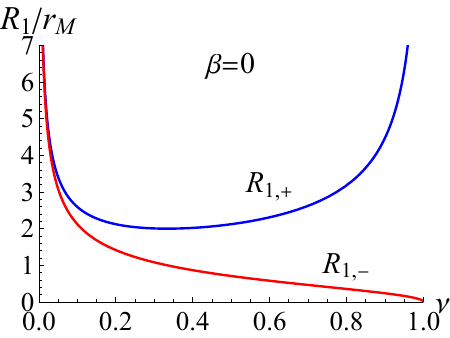}
\includegraphics[width=5.5cm]{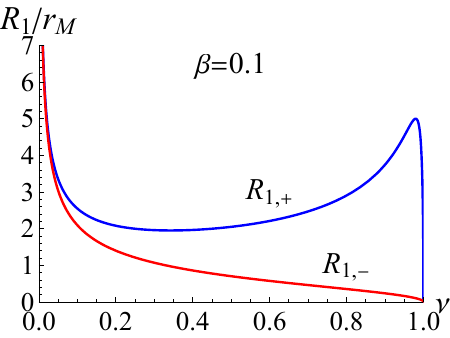}
\includegraphics[width=5.5cm]{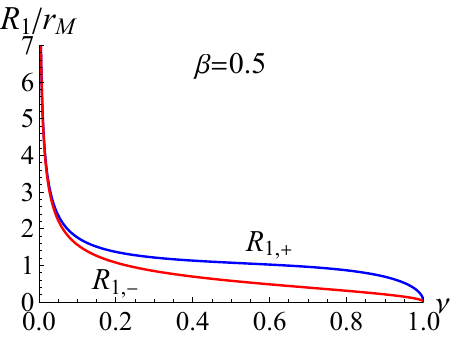}
\caption{$S^1$ radii for the outer and inner rims of the horizon, $R_{1,+}$ and $R_{1,-}$, respectively. \label{fig:s1rad}}
\end{figure}

\begin{figure}[t]
\includegraphics[width=6cm]{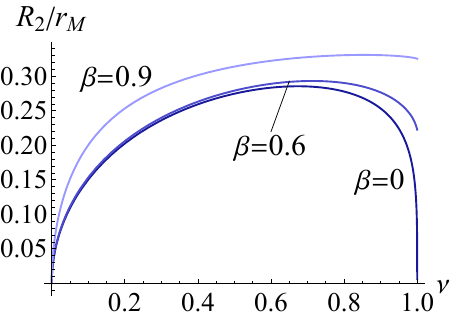}
\caption{$S^2$ radius of the ring horizon. \label{fig:s2rad}}
\end{figure}

Finally, let us consider the shape of the ring horizon. 
As previously introduced in Ref.~\cite{Elvang:2006dd}, it is convenient to define the following three scales that characterize the shape of a black ring:

\begin{itemize}
\item[(i)]
The $S^1$ radii of the outer and inner rims of the event horizon are given by
\begin{align}
R_{1,\pm}  := \sqrt{g_{\psi\psi}}\biggr|_{y=-1/\nu,x=\mp 1} = \frac{2\ell (1+\nu)\sqrt{1+\beta^2+\beta^4}}{1\mp\nu+(1\pm\nu)\beta^2}. \label{eq:def-R1pm}
\end{align}
\item[(ii)]
The $S^2$ radius defined in terms of the area of $S^2$ is give by
\begin{align}
R_2 :=\sqrt{\frac{{\rm Area \ of \ } S^2}{4\pi}}= \sqrt{\fr{2} \int_{-1}^1  \sqrt{g_{\phi\phi} g_{xx}}\bigr|_{y=-1/\nu} dx}.
\end{align}
\end{itemize}
As can be seen from Figs.~\ref{fig:s1rad} and \ref{fig:s2rad}, regardless of the values of the charge $\beta$,
$\nu\to 0$ corresponds to the thin ring limit, as the $S^1$ radii at $\nu\to 0$  diverge as $\nu^{-1/2}$, specifically:
\begin{align}
 \frac{R_{1,+}}{r_M} \simeq \frac{R_{1,-}}{r_M} \simeq \frac{\sqrt{1-\beta^6}}{(1+\beta^2)^{3/2}} \fr{\sqrt{2\nu}}.
\end{align}
Additionally, 
\begin{align}
\frac{R_{1,+}}{r_M}-\frac{R_{1,-}}{r_M} \simeq 
 \frac{(1-\beta^2)\sqrt{1-\beta^6}}{(1+\beta^2)^{5/2}}\sqrt{2\nu},
\end{align}
and 
\begin{align}
\frac{R_2}{r_M} \simeq \fr{2} \sqrt{\frac{\nu}{1-\beta^6}}.
\end{align}
In these limits, the ring thickness defined in terms of the two $S^1$ radii becomes infinitely thin as $\nu^{1/2}$, and similarly, the one defined in terms of the $S^2$ radius also behaves as $\nu^{1/2}$.

\medskip


In contrast to $\nu \to 0$, the limit $\nu \to 1$ results in different shapes for the vacuum black ring ($\beta=0$) and the charged dipole black ring ($\beta \neq 0$).
For $\beta=0$, as discussed in Ref.~\cite{Elvang:2006dd}, the outer radius extends to infinity, but the inner radius shrinks to zero, and the $S^2$-radius also shrinks to zero. Consequently, the vacuum black ring approaches a very thinly flattened disk in this limit.
However, for $\beta \neq 0$, the two $S^1$ radii shrink to zero while maintaining a constant ratio:
\begin{align}
 \frac{R_{1,-}}{R_{1,+}} \to \beta^2\quad ( \nu \to 1 ),
\end{align}
Meanwhile, the $S^2$ radius converges to a nonzero value:
\begin{align}
 \frac{R_2}{r_M} \to \frac{\beta^{3/2}}{\sqrt{2(1+\beta)(1+\beta^2)}}\quad (\nu\to 1).\label{eq:R2-fat}
\end{align}
As a result, the charged dipole black ring approaches a very thin cylinder whose height is estimated by Eq.~(\ref{eq:R2-fat}). This difference is likely caused by the presence of the dipole charge at the $\nu \to 1$ limit.

\section{Comparison with Elvang-Emparan-Figueras black ring}\label{sec:EEF}

The charged dipole black ring solution (EEF solution) discovered earlier by Elvang, Emparan, and Figueras~\cite{Elvang:2004xi}  also possesses mass, two angular momenta, electric, and dipole charges, but only three of these charges are independent, similar to the black ring solution obtained in Section~\ref{sec:sol}. In this section, we compare the obtained black ring solution with the EEF black ring solution. As a result, we find that these two solutions differ except in the vacuum limit.

\medskip

First, one of their clear differences is that the horizon angular velocity, $\omega_\phi$, along the direction of $S^2$, vanishes for the EEF solution but does not for the obtained solution, except in the vacuum case $\beta = 0$, as can be seen from Eqs.~(\ref{eq:omegai}) and (\ref{eq:EEH}). Then, the two solutions do not share the same phase, except in the neutral limit. Moreover, due to the absence of $\omega_\phi$, the EEF solution does not admit the topology of the ergoregion for the case (ii) in Fig.~\ref{fig:ergo}, as the Emparan-Reall black ring solution~\cite{Emparan:2001wn}.

\medskip

Next, let us examine the differences between the two solutions in the phase diagrams. To do so, we normalized the two angular momenta and the dipole charge given in Eq.~(\ref{eq:EEH}) for the EEF black ring by the mass as follows:

\begin{align}
&j_\psi^{EEF} = -\frac{\sqrt{1-\lambda } \sqrt{\mu +1} \left(C_\lambda ( \mu+1) +3
   \beta ^2 C_\mu (\lambda -1)\right)}{\left(\beta ^2+1\right)^{3/2} \sqrt{2-2 \nu }
   (\lambda +\mu )^{3/2}},\\
 &  j_\phi^{EEF}=
   -\frac{\beta  \sqrt{1-\lambda } \sqrt{\mu +1} \left(\beta ^2
  C_\lambda  (\mu +1)+3 C_\mu (\lambda -1)\right)}{\left(\beta ^2+1\right)^{3/2}
   \sqrt{2-2 \nu } (\lambda +\mu )^{3/2}},\\
 & \bar{q}^{EEF}= \frac{C_\mu \sqrt{6(1-\lambda)(\mu +1)}}{(1-\mu ) \sqrt{(1+\beta ^2)(1-\nu)(\lambda +\mu)}},
\end{align}
where
\begin{align}
C_\lambda  = -\sqrt{\frac{\lambda  (\lambda +1) (\lambda -\nu )}{1-\lambda }},\quad  C_\mu = -\sqrt{\frac{(1-\mu ) \mu  (\mu +\nu )}{\mu +1}}.
\end{align}

Here, we have identified the parameter $\alpha$ in Ref.~\cite{Elvang:2004xi} with our $\alpha$ and replaced it with $\beta$.
The parameters $(\mu,\nu,\lambda)$ in the EEF solution are related to each other by the condition for the absence of conical singularities on the inner disc of the ring, given as

 \begin{align}
 \frac{(\lambda +1) (1-\mu )^3}{(\nu +1)^2} = \frac{(1-\lambda ) (\mu +1)^3}{(1-\nu )^2},
 \end{align}
and the condition for the absence of the Dirac-Misner string singularity
 \begin{align}
 \beta^2 C_\lambda(1-\mu)= 3  C_\mu  (\lambda +1).
 \end{align}

\begin{figure}
\begin{minipage}{0.3\columnwidth}
\includegraphics[width=5.5cm]{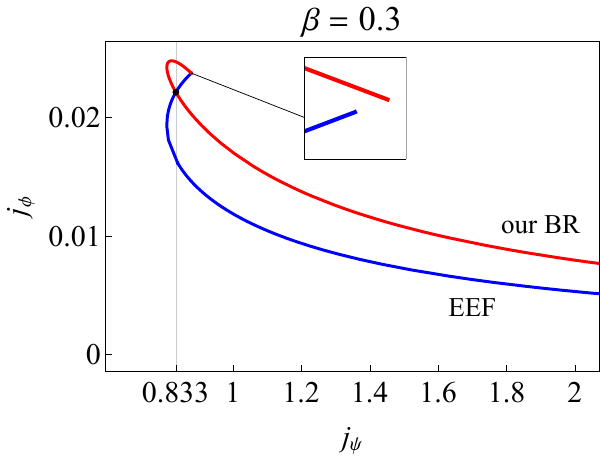}
\includegraphics[width=5.5cm]{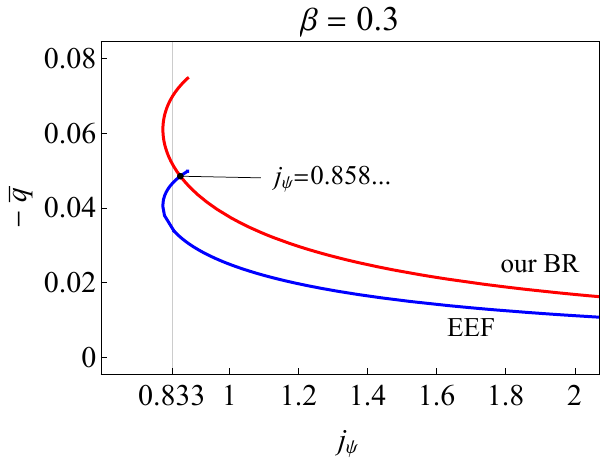}
\end{minipage}
\begin{minipage}{0.3\columnwidth}
\includegraphics[width=5.5cm]{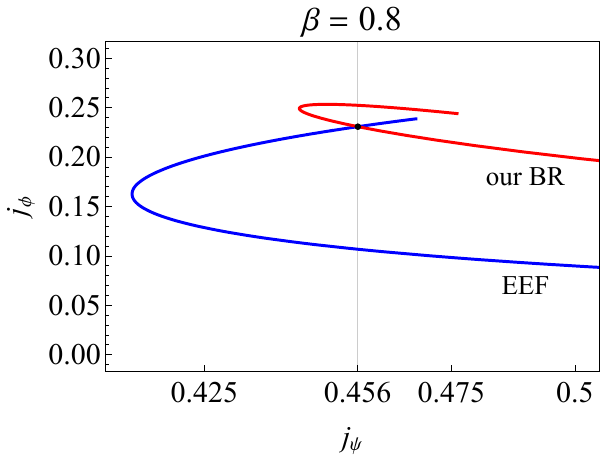}
\includegraphics[width=5.5cm]{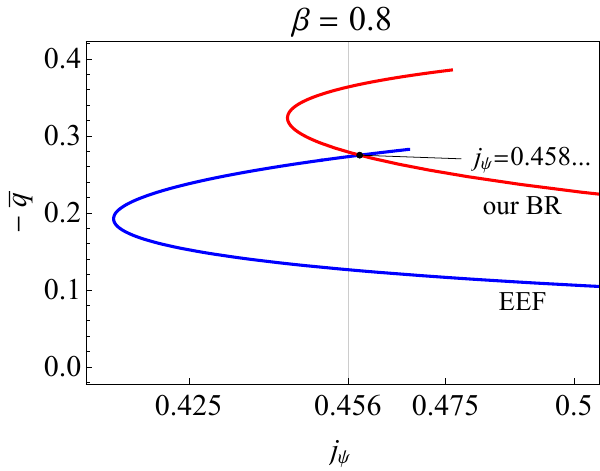}
\end{minipage}
\begin{minipage}{0.3\columnwidth}
\includegraphics[width=5.5cm]{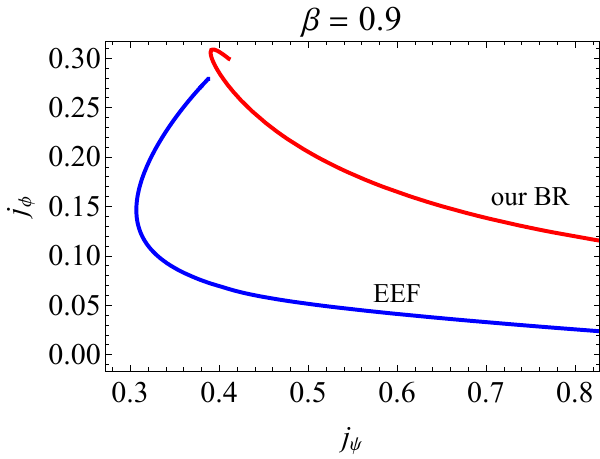}
\includegraphics[width=5.5cm]{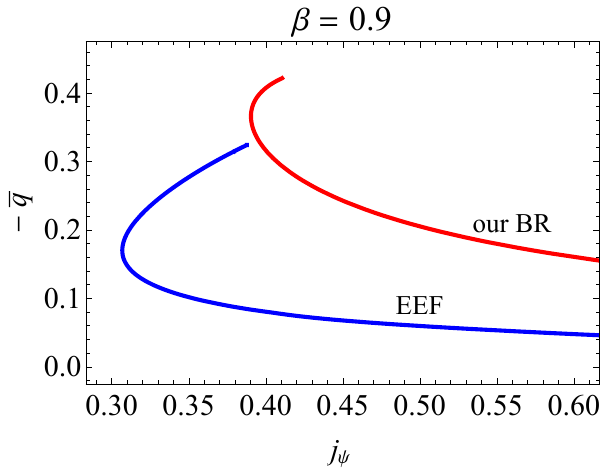}
\end{minipage}
\caption{
Comparison of $\beta={\rm const}$ phases for the obtained solution and EEF solution shown in the $(j_\psi,j_\phi)$ and $(j_\psi,-\bar{q})$ planes. 
For lower charge such as $\beta= 0.3$ and $0.8$, the two solutions can have the phases such that $(j_\psi,j_\phi)$ are equal or $(j_\psi,-\bar{q})$ are equal, but three of them are not equal simultaneously.   
\label{fig:phase-jj}}
\end{figure}

Figure~\ref{fig:phase-jj} illustrates the phases of the obtained black ring and the EEF black ring for $\beta=\tanh\alpha={\rm const.}$  ($\beta=0.3,0.8.0.9$).
We observe that the two solutions at the intersecting point of the curves in each top panel (for $\beta=0.3,0.8$) can have the same four conserved charges: the mass, two angular momenta, and electric charge, but possess different dipole charges, as confirmed in the bottom panel below.

\medskip
Lastly, we also note that the obtained black ring has the vanishing magnetic dipole potential~(\ref{eq:dipolepot}), but  the EEF black ring  admits a nonzero magnetic dipole potential, as expressed by
\begin{align}
\Phi_m = -\frac{\sqrt{3} \pi  c R \mu \sqrt{(1-\lambda)(1+\mu)}  (4 \lambda  \mu +\lambda -3 \mu ) (C_\lambda  (\mu -1)+3 C_\mu   (\lambda +1))}{2 C_\lambda  C_\mu   (\lambda  (8 \mu -1)-9 \mu )}.\label{eq:phi-m-eef}
\end{align}

\section{Summary}\label{sec:sum}

In this paper, we have constructed an exact solution describing a non-BPS charged rotating black ring with a dipole charge in five-dimensional minimal supergravity. This was achieved by employing a combination of two solution-generation techniques: the ISM and the electric Harrison transformation. Furthermore, we investigated various physical properties of the obtained solution, including the ergoregion, phase diagram, and horizon shape.
The resulting solution possesses the mass, two angular momenta, an electric charge, and a dipole charge, although only three of these quantities are independent, similar to another non-BPS charged dipole black ring solution previously found by Elvang, Emparan, and Figueras. We have also delineated the differences between the obtained black ring and the EEF black ring.

\medskip

As mentioned in Ref.~\cite{Elvang:2004xi}, it is conceivable that there exists a most general black ring solution with independent parameters: mass $M$, two angular momenta $J_\psi, J_\phi$, electric charge $Q$, and dipole charge $q$, and seems to have been constructed in Ref.~\cite{Feldman:2014wxa}. 
However, the obstacle in analyzing this solution lies in the considerably lengthy and formal expression. For this reason, we are uncertain whether it truly describes the regular black ring solution with the five independent quantities.
Therefore, to investigate the regularity or physical properties, we need to have a compact form, such as the $C$-metric form, of the solution. 
In our upcoming paper, we plan to explore the construction of such a black ring solution using solution-generation techniques.
Another intriguing avenue for future research involves extending our findings to the realm of charged black holes with multiple horizons. This could include constructing charged counterparts of exotic configurations such as the black saturn~\cite{Elvang:2007rd} or black di-ring~\cite{Iguchi:2007is,Elvang:2007hs}.

\section*{Acknowledgement}
R.S. was supported by JSPS KAKENHI Grant Number JP18K13541. 
S.T. was supported by JSPS KAKENHI Grant Number 21K03560.

\appendix

\section{Relation to the capped black hole solution}
The solution in Ref.~\cite{Suzuki:2023nqf} reduces to the solution~(\ref{eq:metricsol}) 
with the parameter choice
\begin{align}
   b = 0,\quad \gamma = \frac{\nu(3-\nu)}{1+\nu},\quad \tanh^3 \alpha = a,
\end{align}
where the metric functions $H,F,J$ are rescaled by
\begin{align}
 H_{\rm BR}(x,y) := \frac{(1+\nu)^4}{8(1-\nu)^7}H_{\rm cap}(x,y), \quad F_{\rm BR}(x,y) := \frac{(1+\nu)^4}{8(1-\nu)^7}F_{\rm cap}(x,y),\quad
  J_{\rm BR}(x,y) := \frac{(1+\nu)^4}{8(1-\nu)^7}J_{\rm cap}(x,y).
\end{align}

\section{Thermodynamics of Elvang-Emparan-Figueras solution}

Here we recapitulate the thermodynamics of EEF black ring in Ref.~\cite{Elvang:2004xi}.
Since for the first charged dipole black ring solution, a different coordinate orientation is used, we need compare the obtained solution to the EEF solution with $A_\mu \to -A_\mu$. In our convention, the thermodynamic variables are given by
\begin{align}
\begin{split}
&M =\frac{3 \pi  R^2(\mu +1)^2  \left(2 s^2+1\right) (\lambda +\mu )}{4 G_5(1-\nu)},\\
&J_\psi =-\frac{\pi  c \sqrt{1-\lambda } (\mu +1)^{7/2} R^3 \left(c^2 C_\lambda (\mu +1)+3 C_\mu (\lambda -1) s^2\right)}{2   G_5(\nu -1)^2},\\
&J_\phi =  -\frac{\pi  \sqrt{1-\lambda } (\mu +1)^{7/2} R^3 s \left(3 c^2 C_\mu (\lambda -1)+C_\lambda (\mu +1) s^2\right)}{2G_5 (\nu -1)^2},\\
&Q = - \frac{2 M\tanh(2\alpha)}{\sqrt{3}},\\
&A_H = \frac{8 \pi ^2 c (1-\lambda ) (\mu +1)^3 R^3 \sqrt{\mu +\nu }\, |c^2 C_\lambda  (\mu +\nu )+3 C_\mu  s^2 (\lambda -\nu )|}{(\nu -1)^2 (\nu +1) \sqrt{\lambda -\nu }},\\
&\kappa =\frac{\nu  (\nu +1) \sqrt{\lambda -\nu }}{2 c R \sqrt{\mu +\nu } |c^2 C_\lambda  (\mu +\nu )+3 C_\mu  s^2 (\lambda -\nu )|},\\
&\omega_\psi =-\frac{(1-\nu ) (\lambda -\nu ) (\mu +\nu )}{cR \sqrt{1-\lambda } (\mu +1)^{3/2} (1-\nu )R \left(c^2 C_\lambda  (\mu +\nu )+3 C_\mu  s^2 (\lambda -\nu )\right)},\\
&\omega_\phi=0,\\
&\Phi_e = \frac{\sqrt{3} s \left(C_\lambda ^2 (\mu -1) (\mu +\nu )+2 C_\lambda  C_\mu  (\mu -1) (\lambda -\nu )-3 C_\mu ^2 (\lambda +1) (\lambda -\nu )\right)}{c \left(9 C_\mu ^2 (\lambda +1)
   (\lambda -\nu )-C_\lambda ^2 (\mu -1) (\mu +\nu )\right)},\\
&q =\frac{2\sqrt{3} c R C_\mu \sqrt{1-\lambda}(1+\mu)^{3/2}}{(1-\mu)(1-\nu)}, \label{eq:EEH}
\end{split}
\end{align}
where the parameter $\alpha$ in the EEF solution is same as in the new solution, and
\begin{align}
C_\lambda  = \varepsilon \sqrt{\frac{\lambda  (\lambda +1) (\lambda -\nu )}{1-\lambda }},\quad  C_\mu =\varepsilon \sqrt{\frac{(1-\mu ) \mu  (\mu +\nu )}{\mu +1}},\quad \varepsilon=\pm 1.
\end{align}
To compare $J_\psi,J_\phi,q$ in the same signature, we must choose $\varepsilon=-1$.
The parameters $(\alpha,R,\mu,\nu,\lambda)$ can take values for
\begin{align}
R>0,\quad 0<\mu<1,\quad 1<\nu<\lambda,
\end{align}
under the conical free condition
 \begin{align}
 \frac{(\lambda +1) (1-\mu )^3}{(\nu +1)^2} = \frac{(1-\lambda ) (\mu +1)^3}{(1-\nu )^2},
 \end{align}
and the condition for the  absence of a Dirac-Misner string singularity
 \begin{align}
 s^2 C_\lambda(1-\mu)= 3 c^2 C_\mu  (\lambda +1).
 \end{align}
Since the electric charge is given by the same expression as Eq.~(\ref{eq:defQ}), we identify the charge parameter $\alpha$ in both solutions. The EEF solution has a nozero magnetic dipole potential $\Phi_m$, which is calculated from the Smarr formula~(\ref{eq:smarr}) as in Eq.~(\ref{eq:phi-m-eef}).
One can check that in terms of $\Phi_m$, the above thermodynamic variables satisfy the first law~(\ref{eq:1stlaw}).

\end{document}